\documentclass[prd,nofootinbib,preprint,superscriptaddress]{revtex4}
\pdfoutput=1
\usepackage[T1]{fontenc}
\usepackage{amsmath,amssymb}
\usepackage{epsfig}
\usepackage{subfigure}
\usepackage{float}
\usepackage{graphicx}
\usepackage[usenames,dvipsnames]{color}
\usepackage{bigints}
\usepackage{slashed}
\usepackage{multirow}
\usepackage[colorlinks,citecolor=blue]{hyperref}
\usepackage{pdfpages}
\usepackage{color}
\usepackage{comment}
\begin{document}
\title{Concealing Dirac neutrinos from cosmic microwave background}

\author{Anirban Biswas}
\email{anirban.biswas.sinp@gmail.com}
\affiliation{Department of Physics, Sogang University, 
Seoul, 121-742, South Korea}
\affiliation{Center for Quantum Spacetime, Sogang University,
Seoul 121-742, South Korea}
\author{Dilip Kumar Ghosh}
\email{tpdkg@iacs.res.in}
\affiliation{School of Physical Sciences, Indian Association for the Cultivation of Science,
2A $\&$ 2B Raja S.C. Mullick Road, Kolkata 700032, India}
\author{Dibyendu Nanda}
\email{dnanda@kias.re.kr}
\affiliation{School of Physics, Korea Institute
for Advanced Study, Seoul 02455, South Korea}
\begin{abstract}
The existence of prolonged radiation domination prior to
the Big Bang Nucleosynthesis (BBN), starting just after the inflationary
epoch, is not yet established unanimously. If instead, the universe undergoes
a non-standard cosmological phase, it will alter the Hubble expansion rate
significantly and may also generate substantial entropy through non-adiabatic
evolution. This leads to a thumping impact on the properties of relic species
decoupled from the thermal bath before the revival of the standard radiation
domination in the vicinity of the BBN. In this work, considering the Dirac nature
of neutrinos, we have studied decoupling of ultra-relativistic right-handed
neutrinos ($\nu_R$s) in presence of two possible non-standard cosmological phases.
While in both cases we have modified Hubble parameters causing faster expansions
in the early universe, one of the situations predicts a non-adiabatic evolution
and thereby a slower redshift of the photon temperature due to the expansion.
Considering the most general form of the collision term with
Fermi-Dirac distribution and Pauli blocking factors, we have solved
the Boltzmann equation numerically to obtain $\Delta{\rm N}_{\rm eff}$ for the
three right-handed neutrinos. We have found that for a large portion
of parameter space, the combined effect of early decoupling of $\nu_R$
as well as the slower redshift of photon bath can easily hide the
signature of right-handed neutrinos, in spite of precise measurement
of $\Delta{\rm N}_{\rm eff}$, at the next generation CMB experiments
like CMB-S4, SPT-3G etc. This however will not be applicable for the scenarios
with only fast expansion.
\end{abstract}
\maketitle
\section{I\lowercase{ntroduction}}
\label{sec:Intro}
 The anisotropies in leftover radiation from the early universe, known as the cosmic
 microwave background (CMB), is highly sensitive to the presence of extra radiation
 energy at the time of recombination \cite{Jungman:1995bz}. The amount of extra radiation
 energy density is usually parameterized in terms of the effective numbers of
 neutrinos as \cite{Mangano:2005cc} 
\begin{equation}
{\rm N_{eff}}\equiv \frac{\left(\rho_{rad}-\rho_{\gamma}\right)} {\rho_{\nu_L}}
\label{eqn:Neff}
\end{equation} where $\rho_{rad}$ is the total radiation energy density, $\rho_{\gamma}$ is
the energy density of photon and $\rho_{\nu_L}$ is the energy density of a
single active neutrino species. The current data from the measurement of CMB by the Planck satellite \cite{Planck:2018vyg} suggests ${\rm N_{eff}=2.99^{+0.34}_{-0.33}}$ at $95 \%$ CL
(including the baryon acoustic oscillation (BAO) data) which perfectly agrees with the
Standard Model (SM) prediction ${\rm N_{eff}^{SM}}=3.045$
\cite{Mangano:2005cc,Grohs:2015tfy, deSalas:2016ztq}. The next generation CMB experiments
particularly CMB-S4 \cite{Abazajian:2019eic} will be sensitive to a
precision of ${\rm \Delta {N_{eff}}=N_{eff}-N_{eff}^{SM}=0.06}$ at 95$\%$ CL, which
is expected to test all such beyond Standard Model (BSM) scenarios with
light degrees of freedom (DOF) that were in equilibrium with the SM
at some point of the evolution of our universe or produced non-thermally
from the decay or annihilation of other heavy species \cite{Abazajian:2019oqj, FileviezPerez:2019cyn,Nanda:2019nqy,Borah:2020boy,Adshead:2020ekg,Du:2021idh, Luo:2020sho,Luo:2020fdt,Biswas:2021kio,Ganguly:2022ujt,Biswas:2022vkq,Han:2020oet,Li:2021okx,Okada:2022cby, Chen:2015dka}.
In many of such BSM scenarios, the primary motivation
is to explain the tiny nonzero neutrino masses  (see for example \cite{Nanda:2019nqy, Borah:2020boy})
as suggested by neutrino oscillation experiments \cite{T2K:2011ypd,DoubleChooz:2011ymz,DayaBay:2012fng,RENO:2012mkc,MINOS:2013xrl}.
Besides, the nature of neutrinos, whether they are Dirac or Majorana fermion,
is one the most fundamental open questions in particle physics
and there has not been any preference to any of the particular scenario
from the existing experimental data till date. However, from the theoretical
point of view, the Dirac nature essentially demands at least two extremely light
right-chiral components like $\nu_L$s compared to the Majorana case where we
usually requires heavy fermionic DOF for the seesaw mechanisms (see
\cite{Cai:2017mow} for a review). 
Therefore depending on their interactions with the SM particles,
these ultra-relativistic DOF in the early universe may have substantial
contribution to the radiation energy density and hence to the parameter
${\rm N_{eff}}$ that leads to severe constraints on the interactions of $\nu_R$s
with the bath particles \cite{Boehm:2012gr,Kamada:2015era,Zhang:2015wua,Huang:2017egl,
Fradette:2018hhl,Escudero:2018mvt,Jana:2019mez,Depta:2019lbe,EscuderoAbenza:2020cmq}.
For instance, if there are three $\nu_R$s and they were in thermal bath at
the early universe, the Planck 2018 data suggests that $\nu_R$s have to be
decoupled from the SM plasma at temperature higher than 600 MeV\,\,\cite{Xing:2020ijf},
otherwise they will contribute more than the current allowed limit
of ${\rm \Delta {N_{\rm eff}}}\leq 0.285$ at 95\% CL.
Therefore, the decoupling temperature of $\nu_R$ is very crucial as
this will decide $T_{\nu_R}$ at the later epoch which eventually
fixes the contribution to ${\rm N_{eff}}$. Note that if the
neutrino masses are generated only by the standard Higgs mechanism like other
SM fermions, the impact of $\nu_R$s into the parameter ${\rm N_{eff}}$
would be extremely small ($\mathcal{O}(10^{-12})$) \cite{Luo:2020fdt} due to minuscule
Yukawa couplings not allowing $\nu_R$s to attain thermal equilibrium
with the SM bath. Therefore, cosmological probe of the Dirac neutrinos in
the upcoming CMB experiments will confirm new interactions in the neutrino sector.

In \cite{Luo:2020sho}, the authors have introduced new interactions
between $\nu_L$ and $\nu_R$ as effective four-fermion
interactions and set upper limits on such couplings by considering
the impact of new physics in ${\rm \Delta N_{eff}}$. In this article, we show
that the upper bound discussed in \cite{Luo:2020sho}, assuming
the standard radiation dominated era prior to the Big Bang Nucleosynthesis
(BBN), can be significantly relaxed if one alters the cosmological
history of the corresponding epoch \cite{McDonald:1989jd,Kamionkowski:1990ni,Chung:1998rq,Giudice:2000ex,Moroi:1999zb,Allahverdi:2002nb,Allahverdi:2002pu,Acharya:2009zt,Monteux:2015qqa,Davoudiasl:2015vba,DEramo:2017gpl,DEramo:2017ecx,Allahverdi:2010xz,Berlin:2016vnh,Tenkanen:2016jic,Berlin:2016gtr,Starobinsky:1994bd,Dine:1995uk,Allahverdi:2020bys}. Although it has been known
with some precision that the universe was radiation dominated at
the time of BBN \cite{Kawasaki:2000en,Ichikawa:2005vw}, one cannot exclude the
possibility of some component other than radiation dominating the total
energy budget of the universe before BBN. Here we explore this possibility
and discuss the influence of such non-standard cosmological evolution of the
universe on the decoupling of right-handed neutrinos $(\nu_R)$ from
the thermal bath of the SM particles. We consider that the
early universe was dominated by a species whose energy density redshifts with the
cosmic scale factor $a$ as ${\rho_{i}\propto a^{-(4+n)}}$. At first,
we take $n=-1$ which represents an early matter dominated universe \cite{Bezrukov:2009th,Allahverdi:2010xz,Berlin:2016vnh,Tenkanen:2016jic,Berlin:2016gtr,Starobinsky:1994bd,Dine:1995uk} where the energy density is dominated by a non-relativistic
species $M$. In the second case, we consider $n>0$ and that leads to the scenario discussed in \cite{DEramo:2017gpl} where a species $\Phi$, other
than the usual matter and radiation, becomes the dominant component in the
energy density. However, one of the fundamental differences
between the two above mentioned scenarios is that in the case of
early matter domination (i.e. for $n=-1$) the species $M$
should not be absolutely stable like $\Phi$, otherwise it will always remain the 
dominate source in the energy budget since the energy density of $M$,
compared to the radiation, redshifts slowly due to the cosmic expansion.
On the other hand, as the energy density of $\Phi$ falls
faster than the radiation ($\propto a^{-4}$), it will eventually become
sub-dominate at some point of time as the universe expands. Nevertheless, in both the cases,
the expansion rate of our universe, denoted by the Hubble parameter $H$,
increases due to presence of additional source of energy over the radiation. This
gives rise to a faster expanding universe where right-handed neutrinos are decoupled
from the thermal bath at some higher temperature for a given interaction strength. 
The higher decoupling temperature of $\nu_R$ eventually generates a
smaller value of $\Delta{\rm N_{eff}}$.
Moreover, in the first case with $n=-1$, the decay of $M$ into the SM particles
leads to a non-adiabatic expansion with entropy production which results in a
slowly cooling universe compared to the standard case of adiabatic
expansion \cite{Scherrer:1984fd}. This further reduces the ratio $T_{\nu_R}/T$
and hence the contribution of $\nu_R$s in ${\rm N_{eff}}$. Therefore, due to
the combined effect of both the faster expanding universe and
entropy injection in the visible sector, the impact of
$\nu_R$s in the parameter ${\rm N_{eff}}$ gets heavily suppressed compared
to the case with standard ${\Lambda}$CDM cosmology. 

The rest of the paper is organized as follows: in section \ref{sec:model},
we describe the effective four-fermion operators responsible for the
thermalisation of $\nu_R$. In section \ref{sec:Neff:SC}, we have discussed the impact of $\nu_R$ to ${\rm \Delta{N_{eff}}}$ in the standard cosmological scenario while the section \ref{sec:Neff:NSC} is devoted to analyse the effect
of non-standard cosmological histories. Finally, we present our
conclusion in section \ref{sec:concl}. A procedure for simplifying 
the general collision term of the Boltzmann equation has been
presented in Appendix \ref{app:BE}.
\section{F\lowercase{our-fermion interactions of} $\nu_R$}
\label{sec:model}
In the aforementioned discussion, we have stated that the main motivation of this work is to study the impact of non-standard cosmology on the decoupling temperature of right-handed neutrinos $\nu_R$ and its consequences in ${\rm \Delta{N_{eff}}}$. The right-handed neutrinos can be thermalised in the early universe through their interactions with the SM bath. One can write down the following effective four-fermion operators \cite{Luo:2020sho}:
\begin{eqnarray}\nonumber
\mathcal{L} &\supset & G_S \overline{\nu_L}\nu_R \overline{\nu_L}\nu_R + G_S^* \overline{\nu_R}\nu_L \overline{\nu_R}\nu_L + G_P \overline{\nu_L}\nu_R \overline{\nu_R}\nu_L + G_V \overline{\nu_L}\gamma_\mu \nu_L \overline{\nu_R}\gamma_\mu \nu_R \\
&& G_T \overline{\nu_L}\sigma_{\mu\nu} \nu_R \overline{\nu_L}\sigma_{\mu\nu} \nu_R + G_T^* \overline{\nu_R}\sigma_{\mu\nu} \nu_L \overline{\nu_R}\sigma_{\mu\nu} \nu_L,
\end{eqnarray}
where $G_S,\,G_P,\,G_V,\,G_T$ are the effective coupling constants for scalar,
pseudo scalar, vector and tensor type interactions respectively and have dimension
similar to the Fermi constant $G_F$.
Here, we have considered that $\nu_R$
interacts only with the active neutrinos ($\nu_L$). In principle, one should
consider interactions of $\nu_R$ with all other SM particles that were in
thermal bath during decoupling of $\nu_R$ which typically occurred at $T\sim
\mathcal{O}(100)$ MeV. However, for simplicity, in this work we have assumed that
$\nu_R$s have interaction with the left-handed neutrinos only and
based on this assumption we have performed a model independent analysis
in an effective theory framework.
In Table \ref{tab:modMsqr}, we present different processes involved in
the thermalisation of $\nu_R$ and the corresponding amplitude square.
In the next section, we discuss briefly about the contribution of $\nu_R$
in ${\Delta{\rm N_{eff}}}$ within the standard cosmological
evolution of the universe. 

\begin{table*}[htb!]
\begin{ruledtabular}
\begin{tabular}{cccc}
 & possible process & S$\times |{\cal M}|^{2}$ & \tabularnewline
\hline 
 & $\nu_{R}(p_{1})+\nu_{R}(p_{2})\leftrightarrow\nu_{L}(p_{3})+\nu_{L}(p_{4})$ & $8|G_{S}-12G_{T}|^{2}(p_{1}\cdot p_{2})(p_{3}\cdot p_{4})$ & \tabularnewline
 & $\nu_{R}(p_{1})+\overline{\nu_{R}}(p_{2})\leftrightarrow\nu_{L}(p_{3})+\overline{\nu_{L}}(p_{4})$ & $4|{G}_{P}-2G_{V}|^{2}(p_{1}\cdot p_{3})(p_{2}\cdot p_{4})$ & \tabularnewline
 & $\nu_{R}(p_{1})+\nu_{L}(p_{2})\leftrightarrow\nu_{R}(p_{3})+\nu_{L}(p_{4})$ & $4|{G}_{P}-2G_{V}|^{2}(p_{1}\cdot p_{4})(p_{3}\cdot p_{2})$ & \tabularnewline
 & $\nu_{R}(p_{1})+\overline{\nu_{L}}(p_{2})\leftrightarrow\nu_{R}(p_{3})+\overline{\nu_{L}}(p_{4})$ & $4|{G}_{P}-2G_{V}|^{2}(p_{1}\cdot p_{2})(p_{3}\cdot p_{4})$ & \tabularnewline
 & $\nu_{R}(p_{1})+\overline{\nu_{L}}(p_{2})\leftrightarrow\overline{\nu_{R}}(p_{3})+\nu_{L}(p_{4})$ & $16|G_{S}-12G_{T}|^{2}(p_{1}\cdot p_{3})(p_{2}\cdot p_{4})$ & \tabularnewline
\end{tabular}\end{ruledtabular}
\caption{Different processes involved in the thermalisation of $\nu_R$ and the corresponding amplitude square where S represents the symmetry factor corresponding to the matrix elements. }
\label{tab:modMsqr}
\end{table*}
\section{C\lowercase{ontribution to} ${\rm \Delta{N_{eff}}}$  \lowercase{from} $\nu_R$ \lowercase{in the standard cosmology} }
\label{sec:Neff:SC}
As discussed earlier, the Dirac nature of neutrinos requires
the newly added right-chiral parts are as light as the left-handed SM neutrinos.\,\,The
presence of additional ultra-relativistic species in the thermal
plasma at the early universe can give substantial contribution to
the effective relativistic degrees of freedom, ${\rm N_{eff}}$
that can be probed by the CMB experiments. From
Eq.\,\,\eqref{eqn:Neff}, the additional contribution coming
from $\nu_R$ at the time of CMB can be written as 
\begin{eqnarray}
\Delta {\rm N_{eff}} &=& \frac{\sum_\alpha \rho_{\nu_R}^\alpha}{ \rho_{\nu_L}},\nonumber \\ 
&=& 3\times \frac{\rho_{\nu_R}}{\rho_{\nu_L}},\nonumber \\
&=& 3\times \left(\frac{T_{\nu_R}}{T_{\nu_L}} \right)^4,
\label{eqn:Neff2}
\end{eqnarray}
where $\alpha=3$, represents the number of right-handed neutrinos in the theory. In the above equation we have assumed all the three $\nu_R$s behave identically and accordingly $\sum_{\alpha} \rho_{\nu_R} = 3\times \rho_{\nu_R}$, where $\rho_{\nu_R}$ is the energy density of a single right-handed neutrino species. To estimate $\Delta {\rm N_{eff}}$ due to $\nu_R$, we need to know the temperature
$\left( T_{\nu_R}\right)$ of $\nu_R$ at the time of CMB, which evolves independently
after the decoupling of $\nu_R$ from the thermal bath.  The decoupling temperature
$\left({ T_{dec}}\right)$ is usually defined as the temperature when the expansion rate of the universe dominates over the interaction rate $\left({\rm \Gamma}\right)$ and
hence at ${T=T_{dec}}$, 
\begin{equation}
\Gamma \left( T_{dec}\right) = H \left( T_{dec}\right). 
\label{eqn:gamma:H}
\end{equation}
After decoupling from the thermal bath, the energy density of $\nu_R$ redshifts as $a(t)^{-4}$, where $a(t)$ is the cosmic scale factor at any given time $t$. As the SM neutrinos ($\nu_L$) also show the similar behavior due to relativistic decoupling
at $T^{\rm dec}_{\nu_L}\sim 1$ MeV, the ratio $T_{\nu_R}/T_{\nu_L}$ remains unchanged afterwards.\,\,So,
practically we do not need to compute the ratio at the time of CMB, rather
it is sufficient to evaluate the ratio at a much higher temperature $T$
$\left(T > T_{\nu_L}^{\rm dec}\gg T_{\rm CMB} \right)$ when $\nu_L$ shares
same temperature with the photon bath. Accordingly, the Eq.\,\,\eqref{eqn:Neff2}
can also be written as
\begin{eqnarray}
\Delta {\rm N_{eff}} &=& 3\times \left(\frac{T_{\nu_R}}{T}
\right)^4\bigg|_{T> T_{\nu_L}^{\rm dec}}, \nonumber \\
&=& 3 \times \xi^4\bigg|_{T> T_{\nu_L}^{\rm dec}},
\label{eqn:Neff3}
\end{eqnarray}
here $\xi = \dfrac{T_{\nu_R}}{T}$.
Therefore, to evaluate $\Delta {\rm N_{eff}}$ due to $\nu_R$, all we need is $T_{\nu_R}$
at a temperature $T$ just before the decoupling of $\nu_L$.

\begin{figure}[htb!]
\includegraphics[height=8cm,width=10cm]{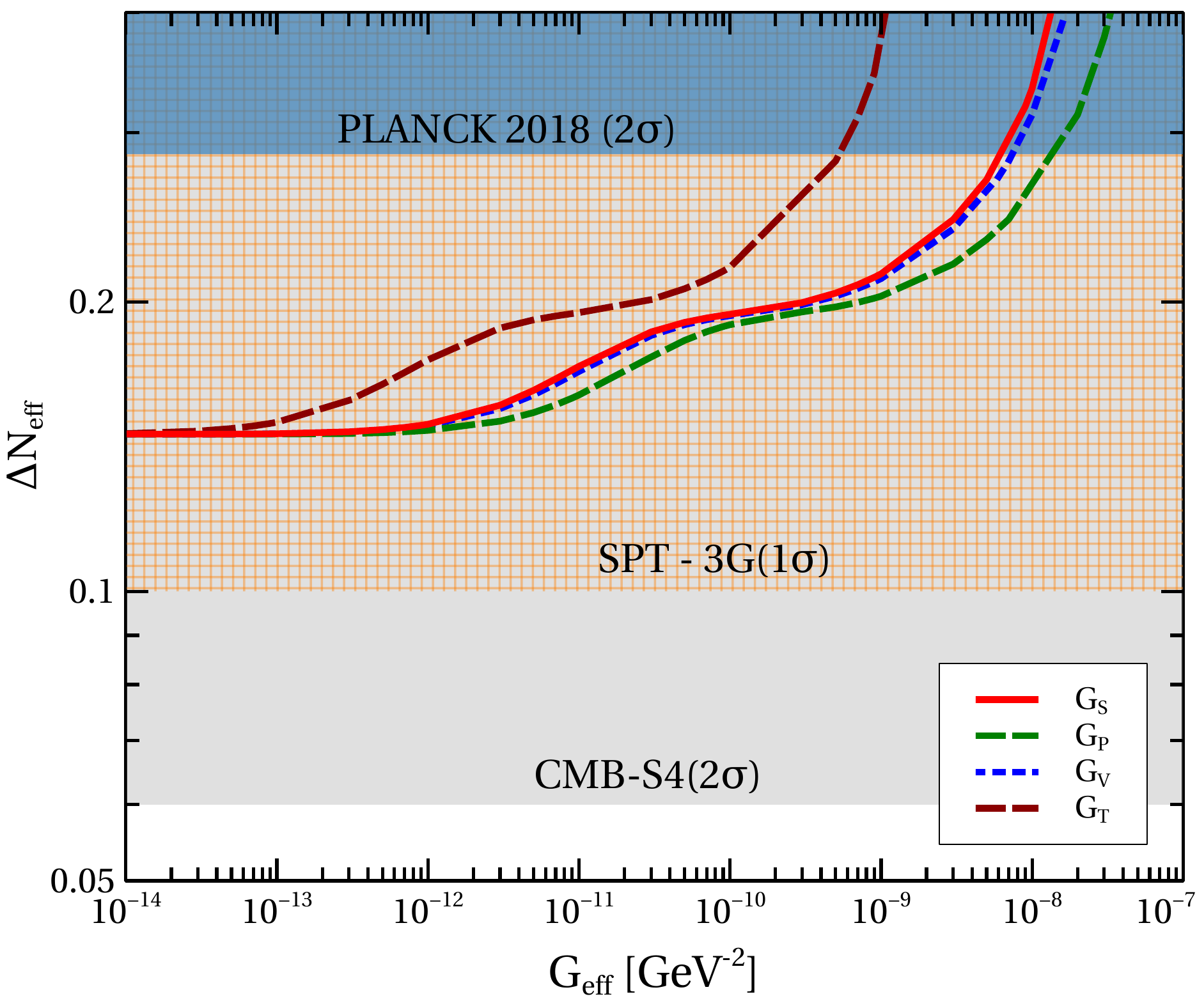}
\caption{The impact of $\nu_R$ to the effective relativistic degrees of freedom
or ${\rm \Delta{N_{eff}}}$ in presence of different interactions (${G_S, G_P,G_V,G_T}$)
as shown in Table \ref{tab:modMsqr}. The present and the future experimental
bounds are also indicated in the same figure.} 
\label{fig:bound:standard}
\end{figure}

In order to proceed further,
we need the Boltzmann equation for the energy density of $\nu_R$ which can also
be expressed in terms of $\xi$ as \cite{Biswas:2021kio}
\begin{eqnarray}
x\dfrac{d\xi}{dx} + \left(\beta -1\right) \xi =
\dfrac{\beta\,x^4}{4\,\kappa\,\xi^3{H}M_0^4} 
\mathcal{C}_{2\rightarrow2}\,,
\label{eq:BE_xi}
\end{eqnarray}
where, $M_0$ is any arbitrary mass scale and $x=M_0/T$. The quantity $\beta$
depends on the variation of DOF ($g_s$) related to the entropy density
with $T$ and its expression is given in 
Appendix \ref{app:BE}. The collision term for $2\rightarrow2$
scatterings listed in Table \ref{tab:modMsqr} is denoted by
$\mathcal{C}_{2\rightarrow2}$. In this work, we have considered the
most general collision term with quantum statistics (FD
distribution) and Pauli blocking factors. To simplify the collision term
we have followed the prescription given in \cite{Hannestad:1995rs, Kreisch:2019yzn}
which reduces the initial twelve dimensional integration into four dimension
and the detailed procedure has been given in Appendix \ref{app:BE}.
We have also checked that our result matches with the one given
in \cite{Luo:2020sho} where the authors had found out the contribution
of $\nu_R$ to $\Delta {\rm N_{eff}}$ by considering the standard
cosmological evolution of our universe and put upper limits
on the effective coupling constants for different types of interactions as shown in Table \ref{tab:modMsqr}.
In Fig.\,\,\ref{fig:bound:standard}, we show the contributions of various
four-fermion operators to ${\rm \Delta{N_{eff}}}$ as a function of
the corresponding effective coupling constant. The most stringent constraint
is coming for the tensor type interactions. In terms of the Fermi constant ${G_F = 1.1664\times10^{-5}\, {\rm GeV}^{-2}}$
 the upper bonds that we obtain can be expressed as,
{\small \begin{equation}
 G_{S}\,<\,5.52 \times 10^{-4}\, G_{F},\, {G}_{P}\,<\,1.28 
 \times 10^{-3} G_{F},\, G_{V}\,<\,6.4 \times 10^{-4}\, G_{F},\, G_{T}\,<\,4.56 
 \times 10^{-5}\, G_{F}.
\label{eqn:upp:lim}  
\end{equation}}
However, as we have mentioned earlier, these upper bounds are true provided
the universe was radiation dominated throughout its evolution starting from
inflation to the matter radiation equality (redshift $z\sim 3400$).
This is not a necessary condition from any cosmological observations
so far and one can always consider some alternative cosmological histories.
In the next section, we will show that these upper limits can be significantly
relaxed if we consider non-standard cosmological history instead of
the standard $\Lambda$CDM cosmology.
Moreover, in Fig.\,\,\ref{fig:bound:standard}
it appears that $\Delta{\rm N}_{\rm eff}$ saturates
to a particular value as we lower the coupling $G_{\rm eff}$
below $10^{-12}$ GeV$^{-2}$. This is mainly due to the reason
that a relativistic species which was in thermal equilibrium in the early universe,
always has a minimum contribution to $\Delta{\rm N}_{\rm eff}$ and
it does not depend on the decoupling temperature ($T_{\rm dec}$). 
For a single species of $\nu_R$, the minimum value is $\Delta{N}_{\rm eff}
\simeq 0.027\times 2\times \dfrac{7}{8}
\left(\dfrac{106.75}{g_s(T_{\rm dec})}\right)^{4/3} = 0.04725$ \cite{Abazajian:2019oqj}.
In this work, we have considered $G_{\rm eff} \geq 10^{-14}$ GeV$^{-2}$
such that the thermalisation condition is always maintained. 
\section{N\lowercase{on-standard cosmological scenarios}}
\label{sec:Neff:NSC}
\subsection{M\lowercase{atter dominated universe}}
\begin{figure}[htb!]
    \centering
    \includegraphics[width=0.8\textwidth]{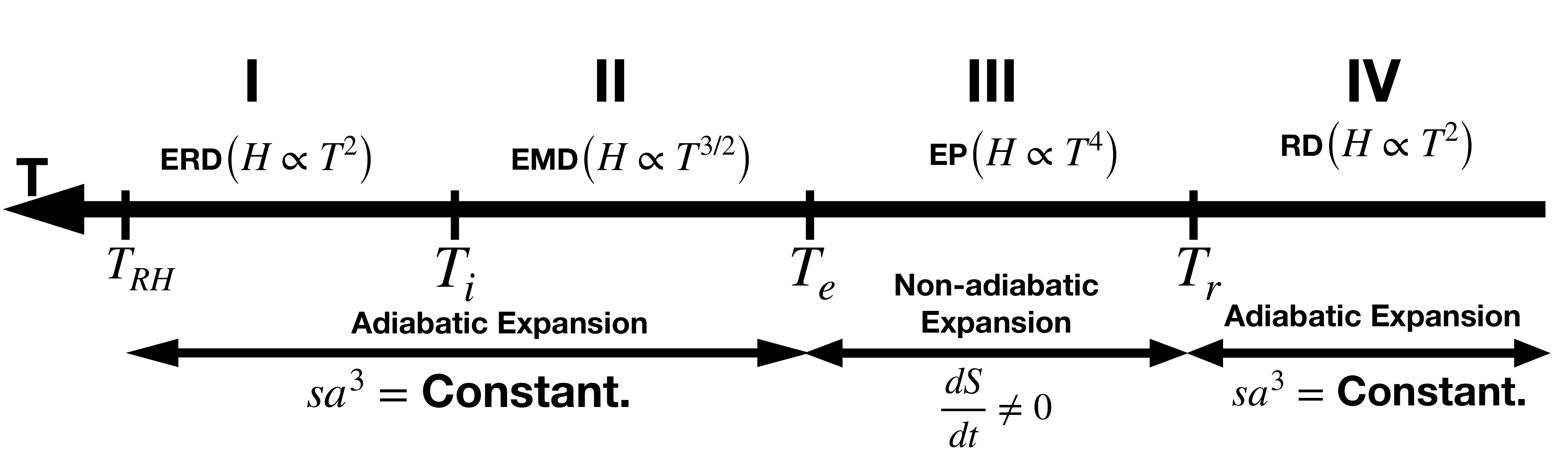}
    \caption{The evolution history of the universe in a early matter dominated universe.}
    \label{fig:history}
\end{figure}

Let us consider that at the early universe just after the inflation, the radiation wasn't the only component that had significantly contributed to the total energy density. Rather, for some epoch, the total energy budget was dominated by some pressureless fluid, denoted as $M$, and the energy density of such species
depends on the cosmic scale factor $a$ like the usual non-relativistic species (referred as matter) as
\begin{equation}
\rho_{M}\propto a^{-3}.
\label{eq:rho:MD}
\end{equation}
In such a scenario, the universe went through different cosmological epochs as shown in Fig.\,\,\ref{fig:history}. After starting with the inflationary epoch of rapid expansion, the universe enters into an early radiation dominated epoch (ERD) at the reheating temperature $T_{\rm RH}$ which is usually defined as the maximum temperature in the radiation dominated universe. The early matter domination (EMD) begins when energy density associated with the species $M$ starts to dominate over that of the radiation at some temperature $(T_i)$. As the rate at which the energy density of matter redshifts is slower than the radiation, therefore unless $M$ decays into the radiation at some later epoch, the matter would always dominate the total energy budget. However, as we know that the universe was radiation dominated (RD) at the time of BBN, the species $M$ must decay to the radiation prior to the formation of light elements. Let us consider that at temperature $T_e$ the decay becomes effective and this results in an enhancement in $\rho_R$ and the radiation again starts to dominate the universe's energy budget at $T_r$ after $M$ decays completely\footnote{Here, we will study the effect of EMD
and EP eras on the decoupling temperature of $\nu_R$, which also depends
on the coupling $G_{\rm eff}$. To understand the maximum possible impact
of non-standard cosmology, we have set $G_{\rm eff}$ in such a way that
$\nu_R$ always decouples between $T_i$ and $T_r$ when, the total energy
density is dominated by the species $M$.}. The cosmological era between $T_e$ and $T_r$ when the species $M$ decays completely into the radiation and creates entropy in the SM bath is known as the entropy production era (EP). The phenomenological consequences of such non-standard cosmological history
have been studied in many different contexts \cite{Vilenkin:1982wt,Starobinsky:1994bd,Dine:1995uk,Allahverdi:2010xz,Berlin:2016vnh,Tenkanen:2016jic,Berlin:2016gtr,Evans:2019jcs,Cosme:2020mck}.

Now, we shall describe briefly the three non-standard epochs occurred before the usual radiation domination in the vicinity of BBN. The basic difference of the EP era with others is that during this particular epoch the universe undergoes a non-adiabatic expansion and as a result the entropy per co-moving volume $S = s\,a^3$ is not conserved. When there is both matter as well as radiation and both have non-negligible
contributions in the energy density ($\rho$), the Hubble parameter is given by $H=\dfrac{1}{M_{Pl}}
\sqrt{\dfrac{8\pi}{3}\left(\rho_R + \rho_M\right)}$ with $M_{Pl}=1.22\times10^{19}$ GeV, the value of the Planck mass. In the ERD era (region I), $\rho_R >> \rho_M$ and $\rho \simeq \rho_R = \dfrac{\pi^2}{30}g_\rho T^4$, where $g_{\rho}$ is known as the number of effective relativistic degrees freedom associated with the radiation energy density. Therefore the expansion rate is same as the usual radiation dominated era i.e.
\begin{eqnarray}
H_{\rm ERD}(T) = 
\sqrt{\dfrac{4\pi^3}{45}g_{\rho}(T)}\dfrac{T^2}{M_{Pl}}\,\,,
\,\, \text{for}\,\, T \geq T_i
\label{eq:Hubble_ERD}
\end{eqnarray}    
 At $T=T_i$,
energy density of the species $M$ becomes equal to the radiation density i.e.\,$\rho^{i}_{M} = \rho_R(T_i)$ and thereafter the EMD era (region II) starts. As the entropy per comoving volume is  conserved before $T \geq T_e$, we can write the energy density of $M$ in any arbitrary temperature ($T \geq T_e$) as
\begin{eqnarray}
\rho_M(T) &=& \rho_M(T_i) \dfrac{g_s(T)}{g_s(T_i)} \dfrac{T^3}{T^3_i}\,,
\nonumber \\
&=& \dfrac{\pi^2}{30} g_{\rho}(T_i) T^4_i\,    
\dfrac{g_s(T)}{g_s(T_i)} \dfrac{T^3}{T^3_i} \,,
\label{eq:rhoM_T}
\end{eqnarray}
wher $g_{s}$ is known as the number of effective relativistic degrees freedom associated with the entropy density of the universe. Using $\rho_M(T)$ one can easily write the Hubble parameter in the EMD era between $T_i$ and $T_e$ as
\begin{eqnarray}
H_{\rm EDM}(T) &=& \dfrac{1}{M_{Pl}}\sqrt{\dfrac{4\pi^3}{45}\,
g_{\rho}(T_i)\,T_i^4}
\sqrt{\dfrac{g_s(T)}{g_s(T_i)} \dfrac{T^3}{T^3_i}}\,,\nonumber\\
&=& H_{\rm ERD}(T_i) \sqrt{\dfrac{g_s(T)}{g_s(T_i)}}
\left(\dfrac{T}{T_i}\right)^{3/2},
\,\,\,\,\text{for}\,\,\,T_i \geq T \geq T_e\,.
\label{eq:Hubble_EMD}
\end{eqnarray}
Comparing Eqs.\,\eqref{eq:Hubble_ERD} and \eqref{eq:Hubble_EMD}
we can easily notice the different temperature dependence of
$H$ in radiation and matter dominated eras. In the ERD epoch,
$H\propto T^2$ while matter domination with adiabatic
expansion leads to $H\propto T^{3/2}$.

Region III in Fig.\,\ref{fig:history} between $T_e$ and $T_r$
is the epoch where decay of the species $M$ into radiation happens
thereby getting back the usual radiation domination at the end. 
Since the energy is injected from the species $M$ to the SM by
the decay, the universe undergoes a non-adiabatic expansion with
entropy production. The evolution of matter density ($\rho_M$),
radiation density ($\rho_R$) and entropy per comoving volume ($S$) 
in this era are given by,
\begin{eqnarray}
&&\dfrac{d\rho_M}{dt} + 3 \rho_M H = -\Gamma_M \rho_M\,, 
\label{eq:rhoM}\\
&&\dfrac{d\rho_R}{dt} + 4 \rho_R H = \Gamma_M \rho_M \,,
\label{eq:rhoR}\\
&& \hspace{-6cm}{\rm and} \nonumber \\
&&\dfrac{dS}{dt} = \Gamma_M\dfrac{\rho_M a^3}{T}\,.
\label{eq:S}
\end{eqnarray}
Where $\Gamma_M$ is the total decay width of $M$. One can easily notice
that when $\Gamma_M << H$, we recover the usual properties of adiabatic
expansion. The solutions of Eqs.\,\eqref{eq:rhoM} and \eqref{eq:rhoR} are 
\begin{eqnarray}
\rho_M &=& \rho_M(T_e) \left(\dfrac{a_e}{a}\right)^3 e^{-\Gamma_M (t-t_e)}\,,
\label{eq:rhoM_sol}\\
\rho_R &=& \rho_R(T_e) \left(\dfrac{a_e}{a}\right)^4 + \dfrac{2}{5}
\dfrac{\Gamma_M}{H(T_e)} \rho_M(T_e) \left(\dfrac{a_e}{a}\right)^{3/2}
\left(1-\left(\dfrac{a_e}{a}\right)^{5/2}\right)\,.
\label{eq:rhoR_sol}
\end{eqnarray}
Where $T_e$ is the initial temperature and the corresponding time and scale factor are
$t_e$ and $a_e$ respectively. The first term of Eq.\,\,\eqref{eq:rhoR_sol} is the usual evolution of $\rho_R$
due to expansion while the second term has the origin from the decay of $M$. Although,
the species $M$ decays into the radiation during this epoch, we still have the matter
dominance in total energy density and using the relation $a \propto t^{2/3}$ one
can easily solve $\rho_R$ in terms of time $t$ as
\begin{eqnarray}
\rho_R &=& \rho_R(t_e) \left(\dfrac{t_e}{t}\right)^{8/3} + \dfrac{3}{5}
{\Gamma_M}\rho_M(t_e) \dfrac{t^2_e}{t}
\left(1-\left(\dfrac{t_e}{t}\right)^{5/3}\right)\,.
\label{eq:rhpR_sol_t}
\end{eqnarray}
From the above expression of $\rho_R$, it is clearly seen that, the first 
term dies out more quickly compared to the term coming from the decay of $M$ with respect to the scale factor $a$. Therefore, within the EP era for $t\gtrsim t_e \left(\dfrac{5}{3} 
\dfrac{t_e}{\tau_{M}}\right)^{3/5}$, we will eventually have a situation when
the second term takes over the first term. As this term has different scale factor
dependence compared to the first one, we have different temperature-scale factor
relationship in the EP era between $t\gtrsim t_e \left(\dfrac{5}{3} 
\dfrac{t_e}{\tau_{M}}\right)^{3/5}$ to $t\simeq \tau_M$,
which is given by (in the limit $a>>a_e$)
\begin{eqnarray}
T \simeq \left(\dfrac{12}{\pi^2}
\dfrac{\Gamma_M}{H(T_e)}\rho_M(T_e)a^{3/2}_{e}\right)^{1/4}
g_{\rho}^{-1/4} a^{-3/8}\,.
\label{eq:T-vs-ainEP}
\end{eqnarray}
The physical meaning of $T\propto a^{-3/8}$
rather than $T\propto a^{-1}$ is that during
ER era, due to energy injection, the temperature
of the universe redshifts slowly with the expansion.
As a result, the entropy per comoving volume,
$S \propto g_s\,a^{3} T^3 \propto g_s\,g^{-3/4}_{\rho}\,a^{15/8}$, 
increases\footnote{during the adiabatic expansion,
$T \propto g^{-1/3}_s \, a^{-1}$ and hence $S$ does
not have any $a$ (or $T$) dependence.}
with the cosmic scale factor. The corresponding
temperature-time relationship can easily be obtained
from Eq.\,\eqref{eq:rhpR_sol_t} as
\begin{eqnarray}
T \simeq \left(\dfrac{18}{\pi^2}
{\Gamma_M}\rho_M(T_e)t^{2}_{e}\right)^{1/4}
g_{\rho}^{-1/4} t^{-1/4}\,.
\label{eq:T-vs-tinEP}
\end{eqnarray}
Therefore, the differential form of temperature-time relationship during
the phase of non-adiabatic expansion is
\begin{eqnarray}
&&\dfrac{dT}{dt} = -\dfrac{HT}{\gamma} \,, \\
&&\hspace{-6cm}{\rm where} \nonumber \\
&&\gamma = \dfrac{8}{3}\left(1+\dfrac{1}{4}\dfrac{T}{g_\rho}
\dfrac{dg_\rho}{dT}\right)\,.
\end{eqnarray}
The Hubble parameter in the EP era (region III) is given by
\begin{eqnarray}
H_{\rm EP} = \dfrac{1}{M_{Pl}} \sqrt{\dfrac{8\pi}{3}\rho_{M}(T_e)}
\left(\dfrac{a_e}{a}\right)^{3/2}\,.
\end{eqnarray}
Now using $\left(\frac{a_e}{a}\right)^{3/2}
\simeq \frac{\pi^2}{12} \frac{H(T_e)}{\Gamma_M\,\rho_M(T_e)}g_\rho T^4$
from Eq.\,\eqref{eq:T-vs-ainEP}, we get 
\begin{eqnarray}
H_{\rm EP} = \dfrac{H(T_e)^2}{\Gamma_M\,\rho_M(T_e)}
\dfrac{\pi^2}{12}\, g_\rho(T) T^4\,, 
\end{eqnarray}
The decay width of $M$ should be of the order of the Hubble parameter
when maximum decay occurs i.e. $\Gamma_M = \zeta H(T_r)$ and this leads
to the end of the EP era at $T_r$. Here $\zeta =\frac{5}{2}$, a constant which we
fix from continuity of the Hubble parameter across the boundary between
EP and RD era (at $T=T_r$). After a few mathematical simplifications
the Hubble parameter during the EP era is given by
\begin{eqnarray}
H_{\rm EP}(T) = \dfrac{1}{H_{\rm RD}(T_r)}\dfrac{4\pi^3}{45}
g_{\rho}(T) \dfrac{T^4}{M^2_{pl}}\,,
\end{eqnarray}
where $H_{\rm RD}$ is the Hubble parameter in the radiation dominated era (region IV)
which has the following well known form
\begin{eqnarray}
H_{\rm RD} (T) = \sqrt{\dfrac{4\pi^3}{45}
g_{\rho}(T)}\,\dfrac{T^2}{M_{Pl}}\,.
\end{eqnarray}
Moreover, continuity of the Hubble parameter across the boundary between
EDM era and EP era i.e. $H_{\rm EMD}(T_e) = H_{\rm EP}(T_e)$ correlates
the three temperatures $T_i$, $T_e$ and $T_r$ in the following way
\begin{eqnarray}
\dfrac{T_e}{T_r} = 
\left(g_{\rho}(T_r)\dfrac{g_{\rho}(T_i)}{g_s(T_i)}
\dfrac{g_{s}(T_e)}{g^2_{\rho}(T_e)}\dfrac{T_i}{T_r}\right)^{1/5}\,.
\label{eqn:MD-para}
\end{eqnarray}
Therefore, only two among the three temperatures are independent, the
rest can be determined by solving the relation iteratively.

{\bf Entropy production}: Here we would like to discuss in more detail
about the entropy production in the region III with necessary expressions.
From Eq.\,\eqref{eq:S}, it is evident that $S$ is not conserved in
the EP era particularly during the decay of $M$ into radiation. The actual
amount of entropy increment can be found after solving Eq.\,\eqref{eq:S}.
In order to solve the entropy equation let us write it in a more
convenient form by replacing $T$ by $S$. After the replacement Eq.\,\eqref{eq:S}
now takes the following form
\begin{eqnarray}
S^{1/3} \dfrac{dS}{dt} = 
\left(\dfrac{2\pi^2}{45}g_s\right)^{1/3} \Gamma_M
\rho_M a^4\,.
\end{eqnarray}
Substituting $\rho_M$ from Eq.\,\eqref{eq:rhoM_sol},
we can get the fractional change in $S$ after
solving the above equation between $t_e$ and $t_r$
(i.e. from temperature $T_e$ up to $T_r$) as
\cite{Scherrer:1984fd, Bezrukov:2009th}
\begin{eqnarray}
\dfrac{S(t_r)}{S(t_e)} \equiv \dfrac{S_r}{S_e} = \left(1+\dfrac{4}{3}\left(\dfrac{2\pi^2}{45}\right)^{1/3}
\dfrac{\rho_M(T_e)\,a^4_e}{S^{4/3}_e}
\,\mathcal{I}\right)^{3/4},
\label{eq:etropy_ratio}
\end{eqnarray}
where, the enhancement factor $\mathcal{I}$ is given by
\begin{eqnarray}
\mathcal{I} = \Gamma_M \int_{t_e}^{t_r}
g_s^{1/3} \dfrac{a}{a_e} e^{-\Gamma_M(t-t_e)}\,dt\,.
\end{eqnarray}
One can simplify $\mathcal{I}$ under certain assumptions that
$t_r >> t_e$, $\Gamma_M\,t_e<<1$ and $g_s$ is not changing
significantly between $t_e$ and $t_r$ then the enhancement
factor has the following simplified expression
\begin{eqnarray}
\mathcal{I} \simeq \Gamma\left[\frac{5}{3}\right]
g_s^{1/3}\left(\dfrac{\tau_M}{t_e}\right)^{2/3}\,.
\end{eqnarray}
Therefore, it is evident that the entropy generation will be
maximum when $M$ has a longer lifetime ($\tau_M$) compared
to $t_e$, the starting point of the region III in Fig.\,\,\ref{fig:history}, or in
other word the universe undergoes a prolonged EP era. Note that the matter domination epoch can be controlled by only two parameters $T_{i}$ and $T_r$ and the other one can be expressed in terms of these two as shown in Eq.\,\,\eqref{eqn:MD-para}. 

\begin{figure}[htb!]
\includegraphics[scale=0.5]{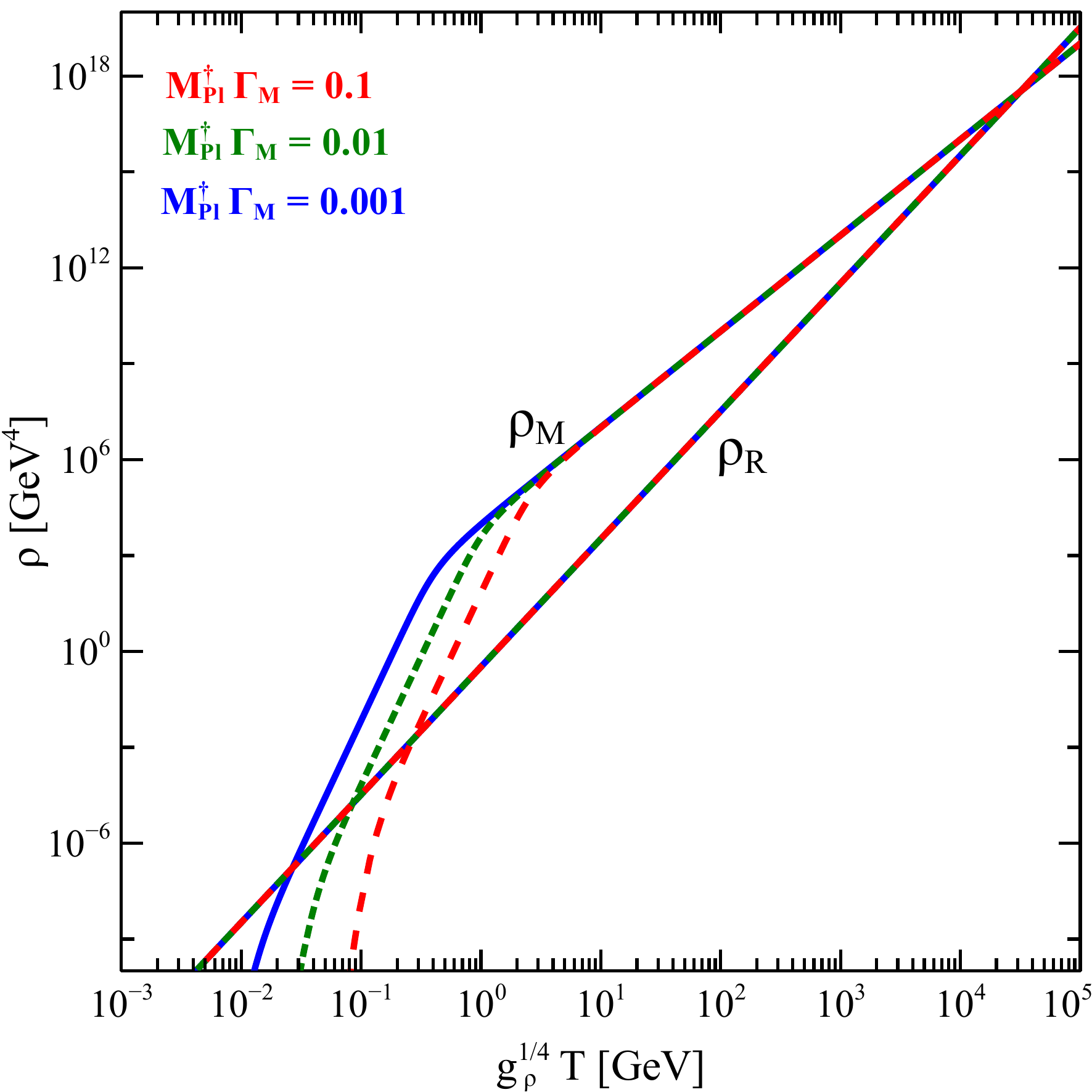}
\caption{Evolution of energy densities of matter and radiation as a function of temperature has shown for three benchmark values of $\Gamma_{M}$ represented in different color and $M^{\dagger}_{Pl}$ is defined as $M^{\dagger}_{Pl} = \sqrt{\frac{3}{8\pi}} M_{Pl}$.} 
\label{fig:energy}
\end{figure}

\begin{figure}[htb!]
\includegraphics[scale=0.53]{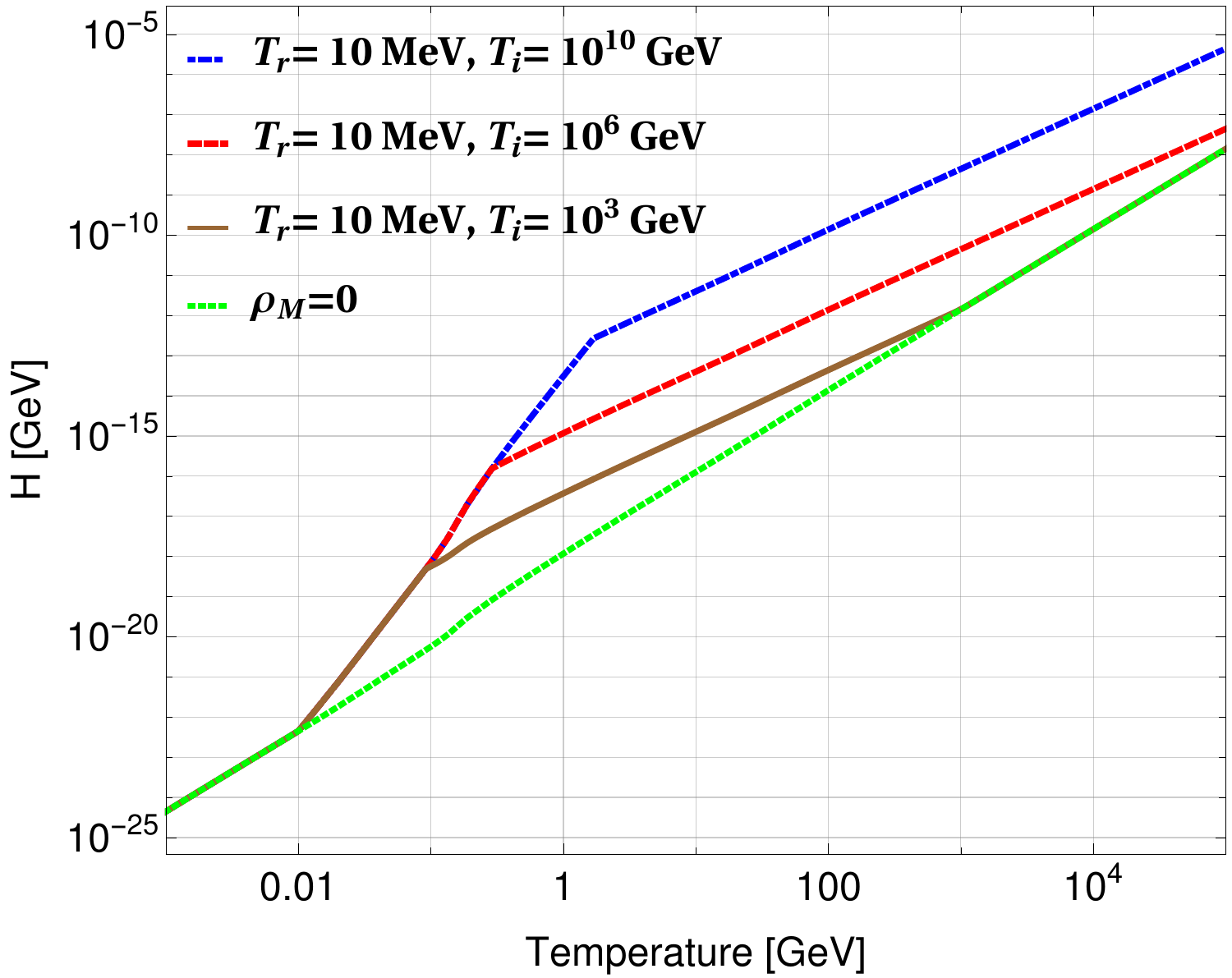}\,
\includegraphics[scale=0.53]{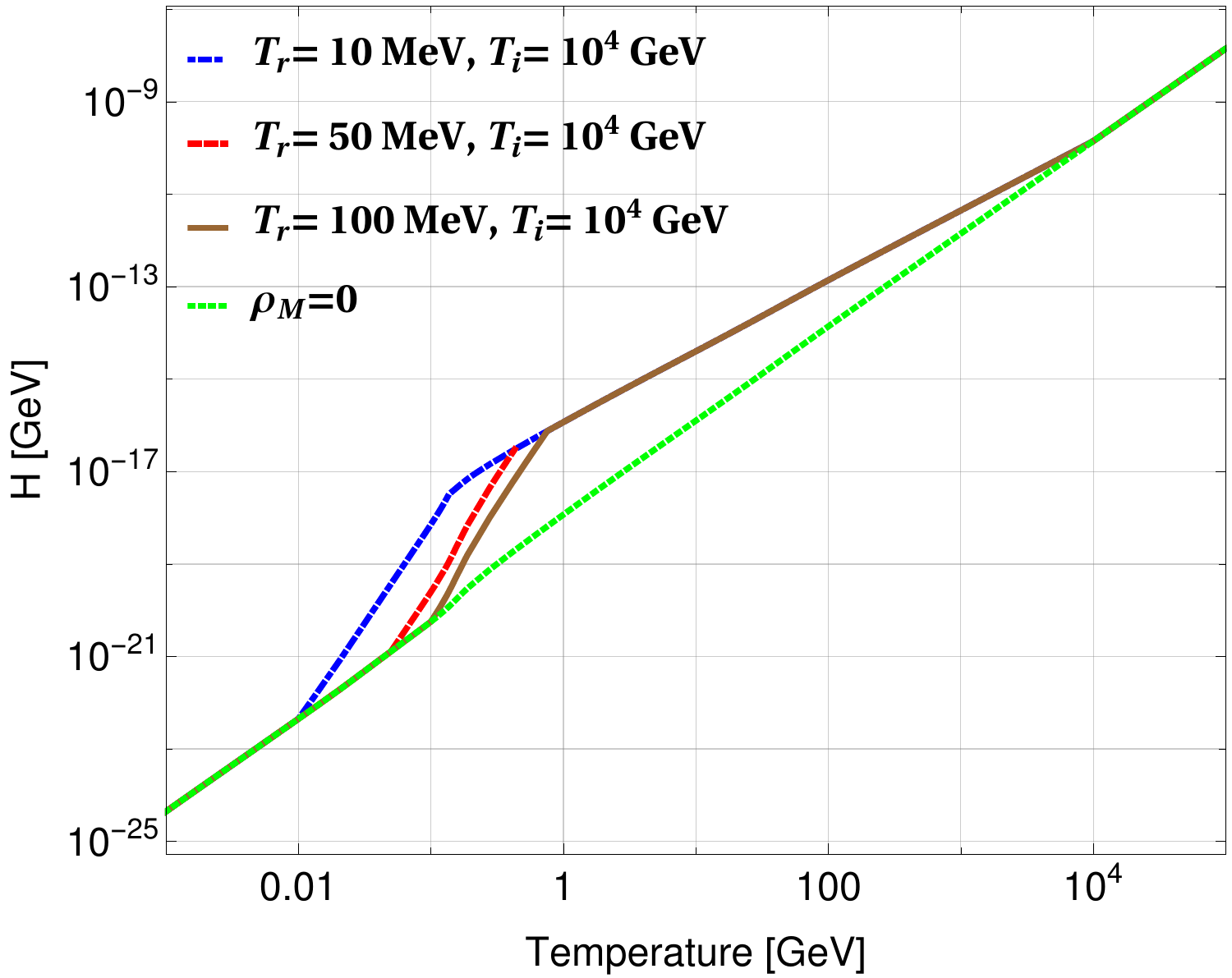}
\caption{The expansion rate for different combinations of $T_i \,\, \text{and}\,\, T_r $ as a function of the bath temperature in the matter dominated universe.}
\label{fig:hubble:MDU}
\end{figure}

\begin{figure}[htb!]
\includegraphics[scale=0.53]{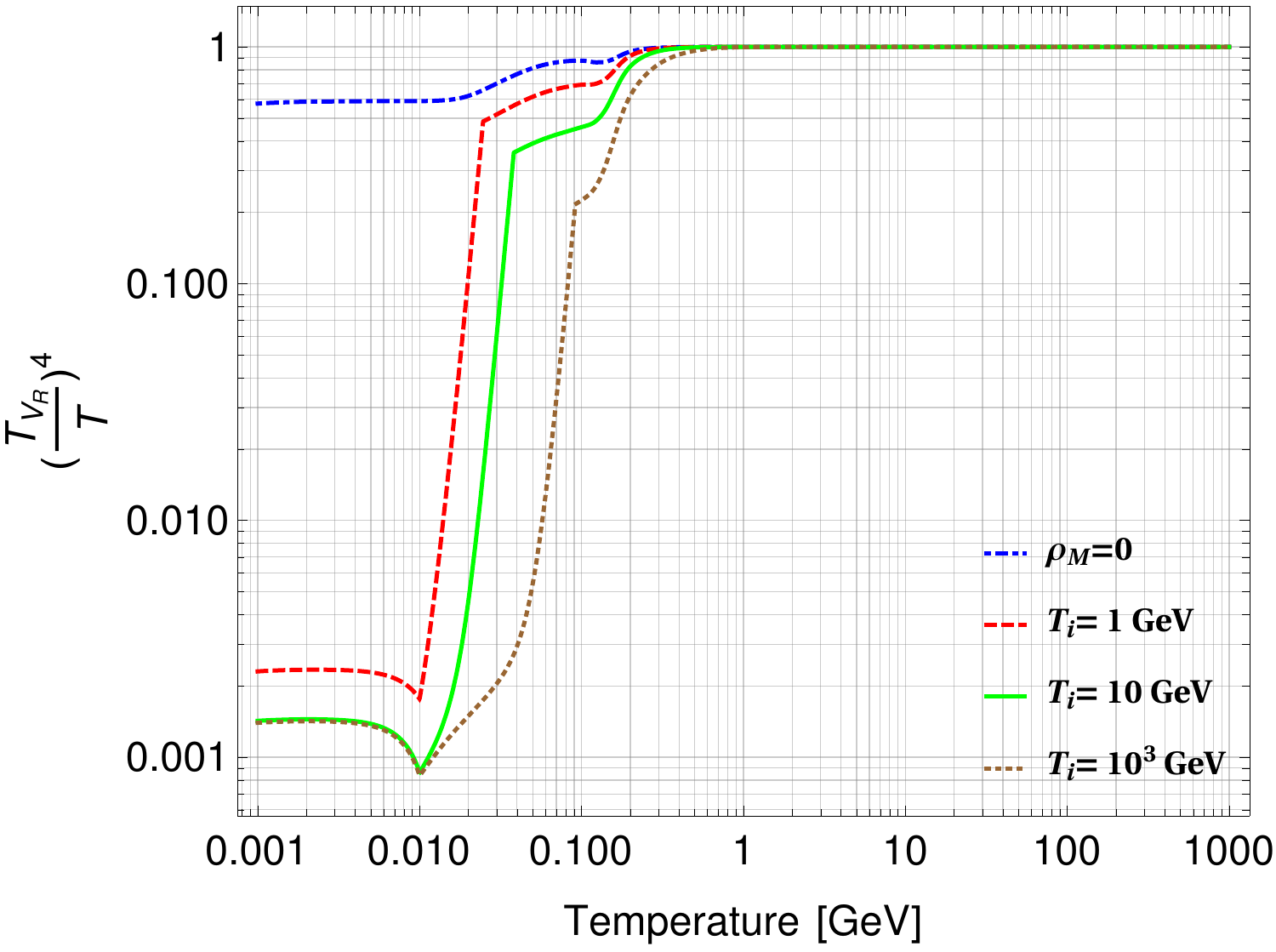}\,
\includegraphics[scale=0.53]{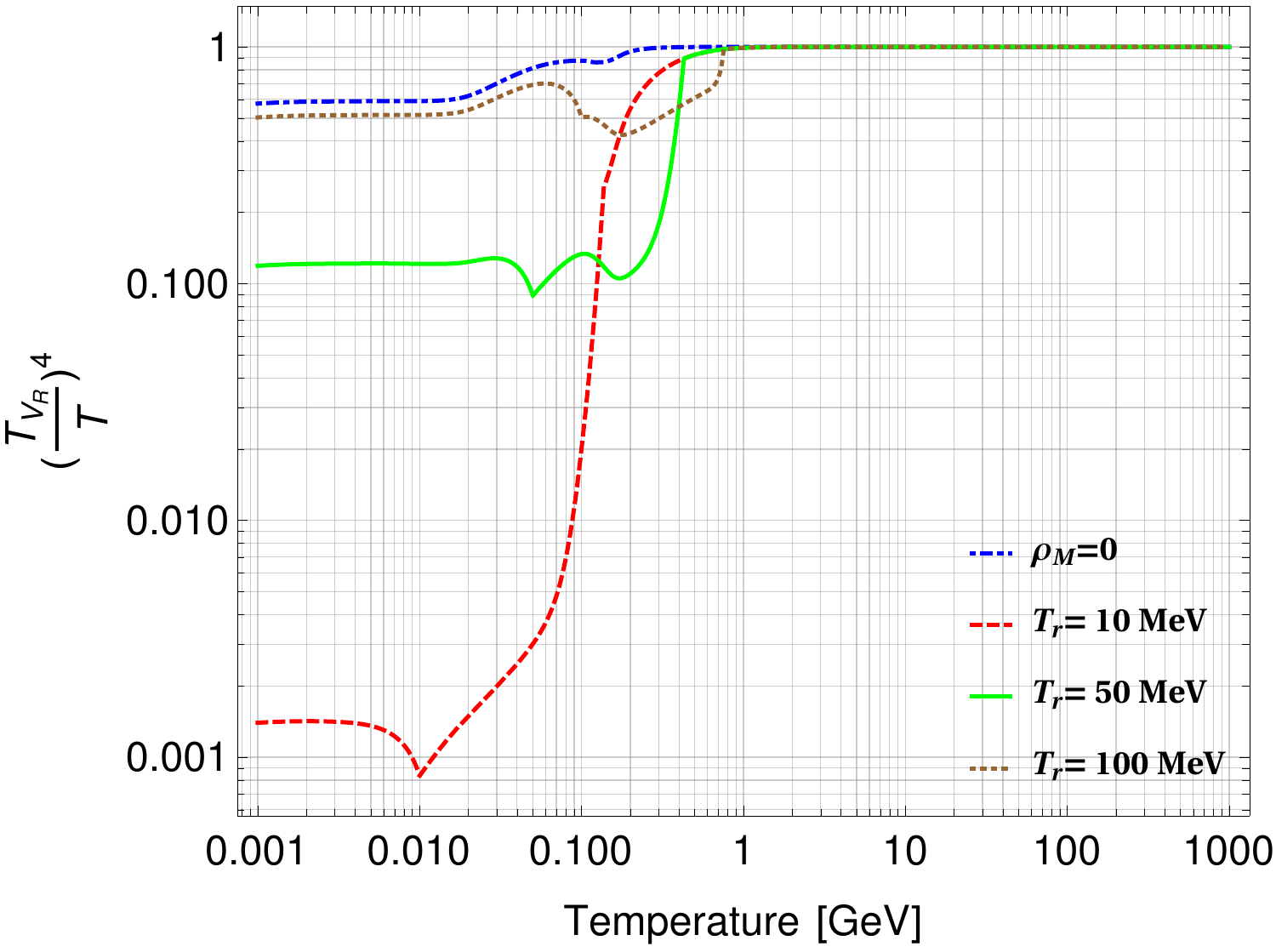}
\caption{Evolution of $T_{\nu_R}$ for different values of the parameters controlling the matter domination for a given ${G_V=10^{-7}\, {\rm GeV}^{-2}}$. In the left panel we have fixed $T_r$ at 10 MeV whereas in the right panel $T_i$ is fixed at $10^4$ GeV. }
\label{fig:dec:MDU} 
\end{figure}

In Fig.\,\,\ref{fig:energy}, we have shown the variation of the energy densities of matter ($\rho_{M}$) and radiation ($\rho_R$) as a function of temperature $(T)$. Both the energy densities decreases as the temperature goes down with the expansion of our universe. The epoch of matter domination solely depends on decay width $\Gamma_{M}$ as shown in Eqs.\,\,\eqref{eq:rhoM} and \eqref{eq:rhoR}. It can be seen that the epoch of matter domination which ends through the entropy injection into the radiation bath becomes longer with the decrease of $\Gamma_{M}$. That means, $T_{r}$, the temperature when $\rho_{M}$ becomes subdominant, decreases with the decrease in $\Gamma_{M}$.        
 In Fig.\,\,\ref{fig:hubble:MDU}, we have shown the impact of the two parameters $(T_i\,\text{and } T_r)$ on the expansion rate of the universe as a function of the temperature. The left panel shows the dependence on $T_i$ (blue for $T_i=10^{10}$ GeV, red for $T_i=10^{6}$ GeV, and brown for $T_i=10^{4}$ GeV) where we have kept $T_r$ to be fixed at 10 MeV whereas the right panel represents the impact of $T_r$(blue for $T_r=10$ MeV, red for $T_r=50$ MeV, and brown for $T_i=100$ MeV) where $T_i$ remains fixed at $10^4$ GeV. The green line in both the figure shows the expansion rate in usual standard radiation dominated universe $\left(\rho_M = 0\right)$. It is clearly seen that in the matter dominated era, the universe expands much faster than the usual radiation dominated universe.  One can also note that, in all these cases, at $T<T_r$ the expansion rate exactly coincide with the standard radiation dominated universe which means that once the matter domination ends we can longer see the impact of the parameters responsible for the non-standard evolution of history. However, as stated before, the non-standard evolution history can affect different cosmological phenomenon and we can always look for their imprints on different cosmological observable and one such observable is
 $\Delta{\rm N_{eff}}$ which can carry the information of such non-standard history. Lets us now discuss consequence of early matter domination era on the $\Delta{\rm N_{eff}}$ coming from the thermalised $\nu_R$ in the early universe. Due to the faster expansion, $\nu_R$s can be decoupled from the thermal bath at some earlier temperature and reduce their final temperature at some later time. However, one important point to note here is that $\Delta{\rm N_{eff}}$ depends on the ratio $T_{\nu_R}/T$ which will not only be affected by the faster expansion of the universe but also the entropy injection in the SM sector from the decay of $M$. The energy density frozen in the non-relativistic matter $M$ had to be transferred to the radiation sector to end the matter domination in order to begin the radiation domination on the on set of BBN.
 Due to this entropy injection the total entropy in the co-moving volume would no longer be constant and the temperature of the thermal bath will start falling slower than the usual as $a^{-3/8}$\cite{Giudice:2000ex, Kolb:2003ke}. The impact of the faster expansion and entropy injection to the SM bath on the evolution of $T_{\nu_R}/T$ is shown in Fig.\,\,\ref{fig:dec:MDU}. The left panel shows the evolution for three benchmark values of $T_i$ by keeping the temperature $T_r$ fixed at 10 MeV whereas in the right panel we have varied $T_r$ by keeping $T_i$ fixed at $10^{4}$  GeV. One can clearly notice that due to the faster expansion the ratio of the temperature drops below 1 at some higher temperature in comparison to the standard radiation domination (the blue dot-dashed line) and then sharply falls due to the entropy injection in the thermal plasma and becomes constant when the entropy injection ends. As a result of this significant drop of $T_{\nu_R}/T$, the value of ${\rm \Delta{N_{eff}}}$ becomes much smaller in comparison to the standard radiation dominated universe. While showing the evolution of the temperature ratio we have considered the vector type interactions only and set ${G_V=10^{-7}\, {\rm GeV}^{-2}}$ whereas all the other kind of interactions has been set to zero. As the early matter domination decreases the final value of $T_{\nu_R}/T$, stronger interactions between $\nu_R$ and SM plasma can be allowed which are excluded in the radiation dominated universe.       

\begin{figure}[htb!]
\includegraphics[scale=0.42]{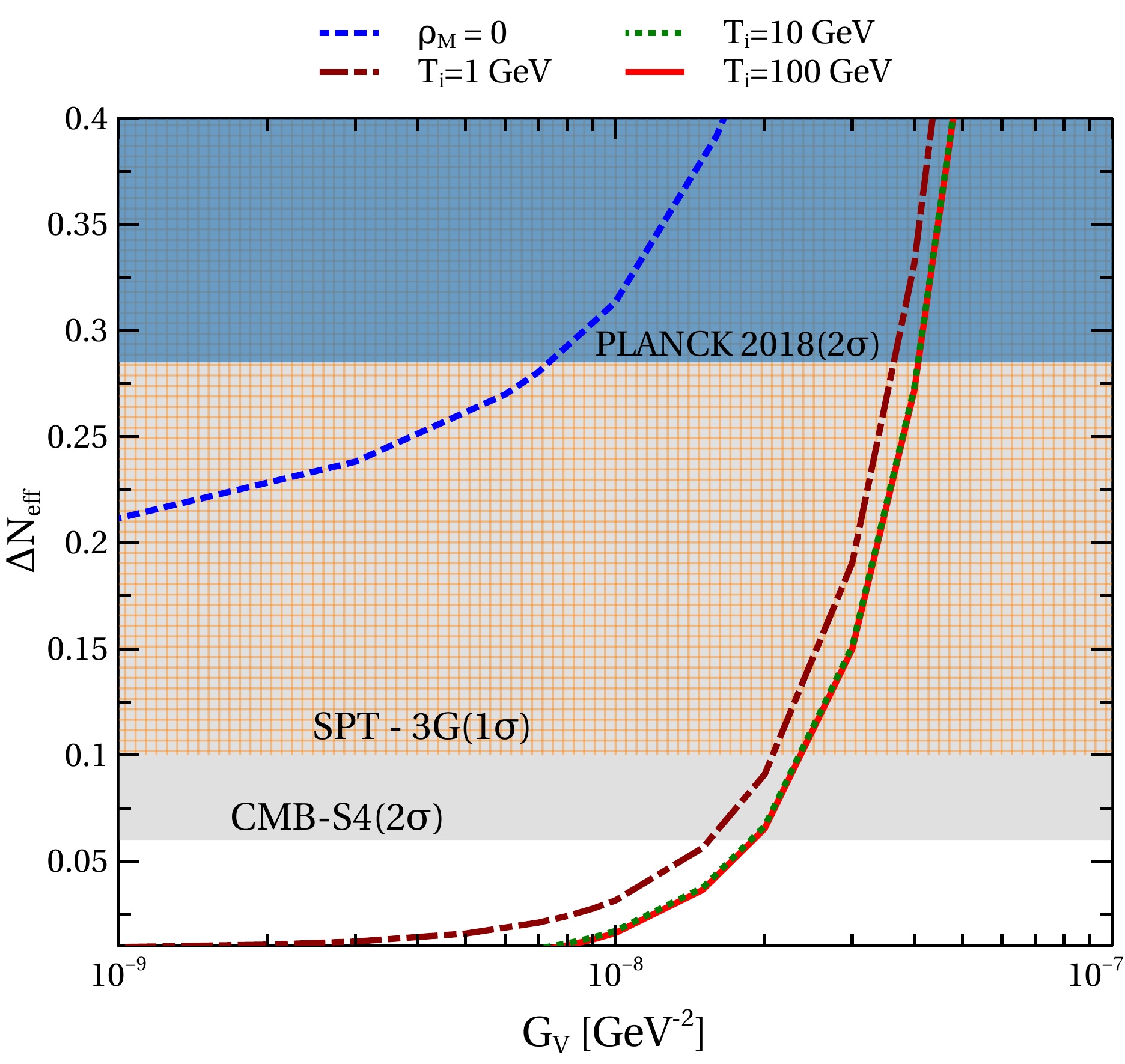}\,
\includegraphics[scale=0.42]{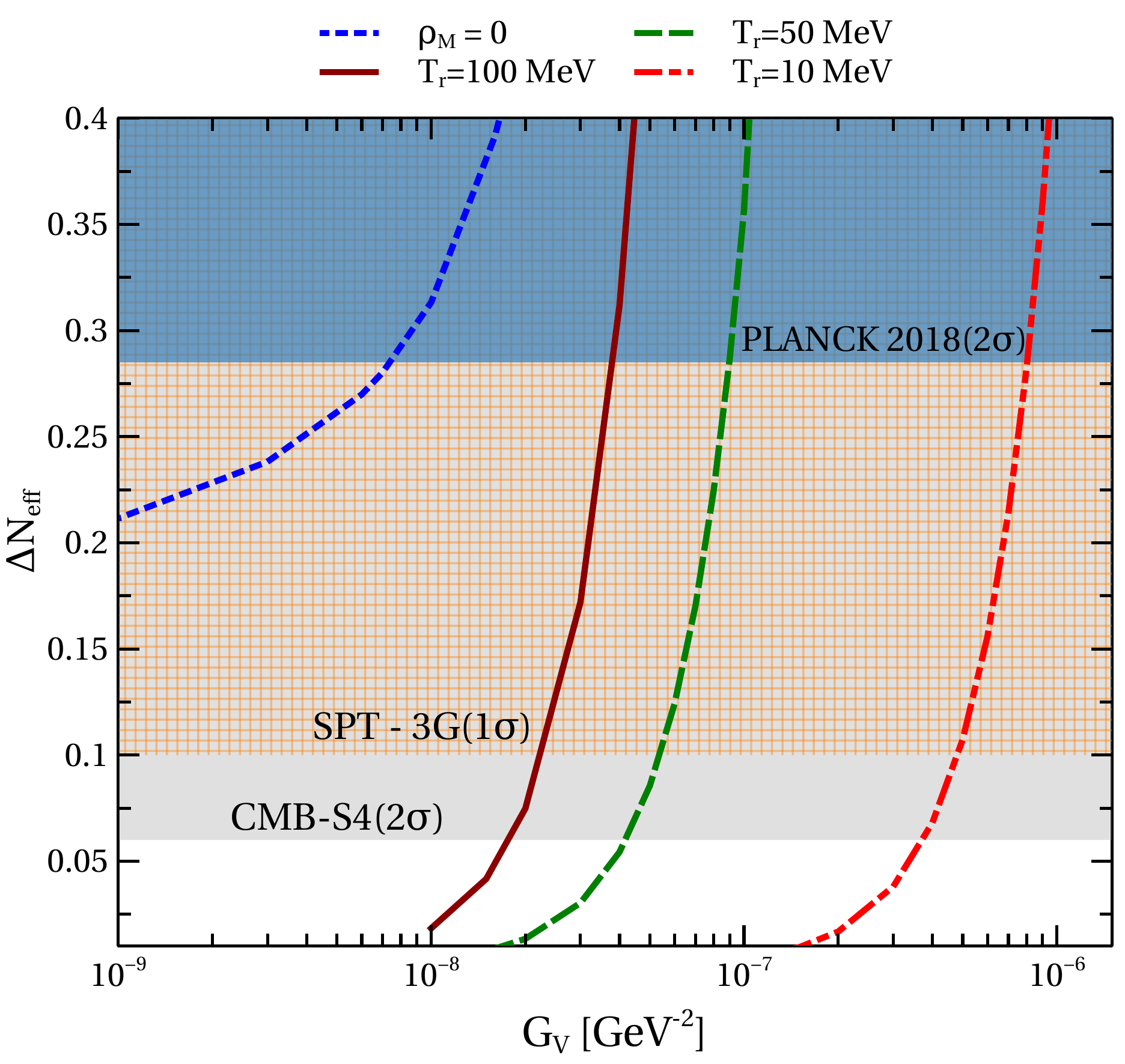}
\caption{The impact of $\nu_R$ to the effective relativistic
degrees of freedom or ${\rm \Delta{N_{eff}}}$
in a early matter dominated universe. Here we have considered
the vectorial interaction with effective coupling $G_V$.} 
\label{scan:MDU}
\end{figure}

Finally, in Fig.\,\,\ref{scan:MDU}, we present the main result of this section. We have shown the variation of $\Delta{\rm N_{eff}}$
as a function of four-fermion interaction strength and here we have considered only the vector type coupling $G_V$.
The other kind of interactions will also show the similar behaviour. As discussed earlier, the faster expansion and entropy injection push
$\Delta{\rm N_{eff}}$ to much smaller value than the standard cosmological scenario for a given interaction strength. In the left panel of Fig.\,\,\ref{scan:MDU}, we show that for a fixed $T_r=$ 100 MeV, $\Delta {\rm N_{eff}}$ decreases with increasing $T_i$. This is because lower value $T_i$ means the lowering the entropy injection to the SM bath. The right panel shows the impact of $T_r$ for three benchmark values where we have kept $T_i$ fixed at $10^4$ GeV. In both the cases, for any given interaction strength, $\Delta{\rm N_{eff}}$ becomes much less than the standard radiation dominated universe (the blue dashed line corresponds to $\rho_M = 0$). In the next section, we have discussion another alternative non-standard cosmological history where the early epoch of the universe was dominated by some species $\Phi$ whose energy density falls even faster than the radiation which can also enhance the expansion rate of the universe and can significantly affect the decoupling temperature of $\nu_R$ and the ${\Delta{\rm N_{eff}}}$.

\subsection{F\lowercase{ast expanding universe}}
Let us consider another non-standard cosmological scenario where the early universe was dominated by a species ($\Phi$) which redshifts faster than radiation,
\begin{equation}
\rho_{\Phi}\propto a^{-(4+n)},
\label{eqn:rho:Phi}
\end{equation}
where $n>0$ for $\Phi$ and $n=0$ for radiation. 
One can express $\rho_{\Phi}$ as a function of the bath temperature ($T$) which can be achieved by incorporating the
conservation of total entropy in a comoving volume ${S=sa^3=\text{constant}}$, where the entropy density(s) can be expressed as,
\begin{equation}
s(T)= \frac{2\pi^2}{45}g_{s} (T) T^3
\end{equation}
As we already know that the universe was radiation dominated at the time of the BBN, let us now define a temperature $T_r$ at which the $\rho_{\Phi}$ becomes equal to $\rho_{R}$ below which the $\rho_{R}$ dominates the total energy budget.
From Eq.\,\,\eqref{eqn:rho:Phi}, one can write

\begin{align}
{\rho_{\Phi}(T)}& = {\rho_{\Phi}(T_r)}\left(\frac{a(T_r)}{a(T)}\right)^{4+n}\\
&= {\rho_{\Phi}(T_r)} \left(\frac{g_{s}(T)}{g_{s}(T_r)} \right)^{\frac{4+n}{3}} \left(\frac{T}{T_r}\right)^{4+n}
\end{align}
where in the last line we use the entropy conservation condition at temperature T and $T_r$. Finally, the total energy can be expressed as 

\begin{equation}
\rho(T) = \rho_{R}(T) + \rho_{\Phi}(T)=\rho_{R}(T)\left[1+\frac{g_{\rho}(T_r)}{g_{\rho}(T)}\left(\frac{g_{s}(T)}{g_{s}(T_r)}\right)^{\frac{4+n}{3}} \left(\frac{T}{T_r}\right)^n\right]
\end{equation}
by setting $\rho_{\Phi}(T_r)=\rho_{R}(T_r)$ as mentioned above. The expansion rate of the universe at the $\Phi$ dominated era (at $T>T_r$) can be controlled by two different parameters $T_r \,\, \text{and}\,\, n $. We use this expression to evaluate the Hubble parameter as,
\begin{equation}
H(T)=\frac{1}{M_{Pl}} \sqrt{\frac{8\pi \rho(T)}{3}}
\label{eqn:hubble:FE}
\end{equation}

\begin{figure}[htb!]
\includegraphics[scale=0.53]{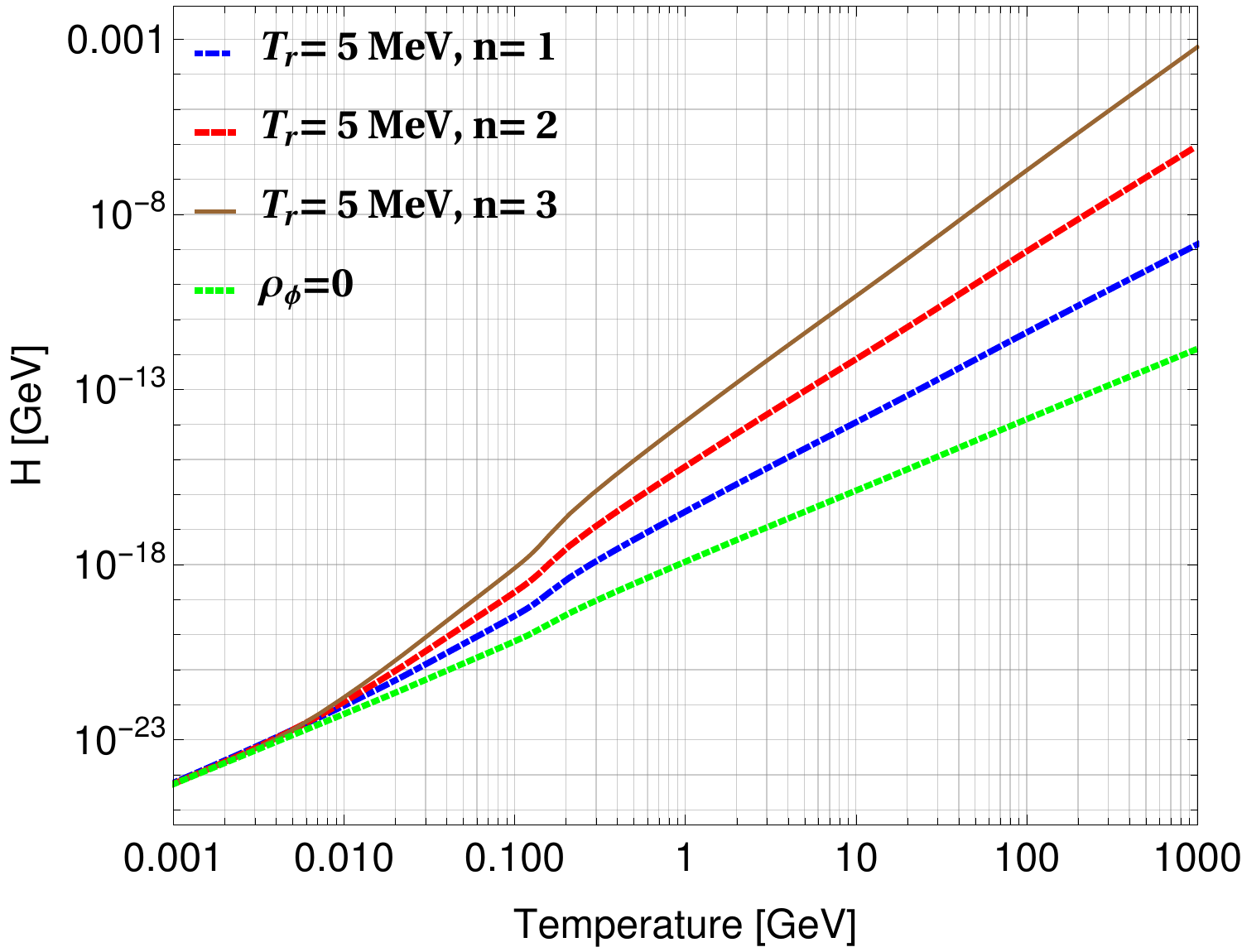}\,
\includegraphics[scale=0.53]{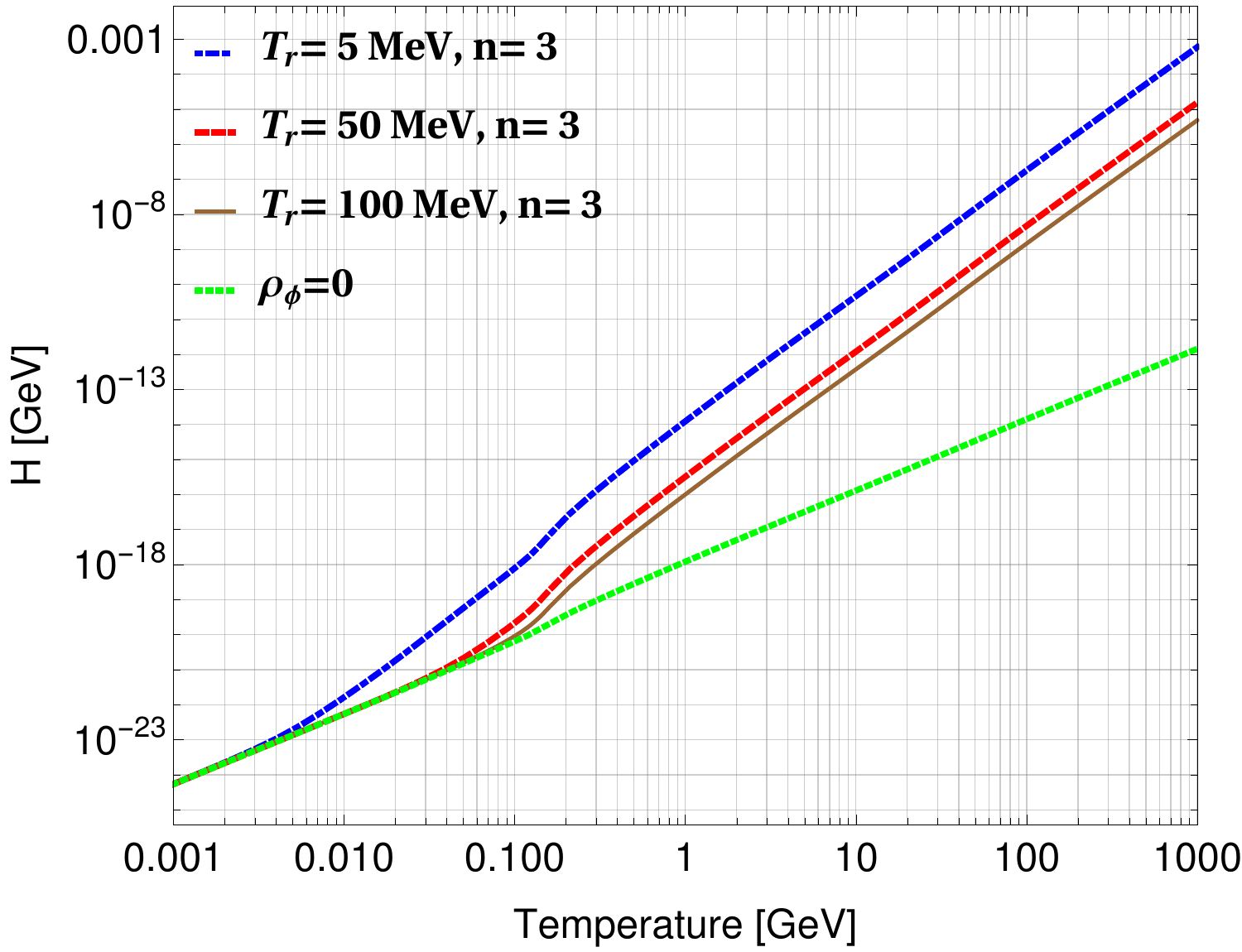}
\caption{The expansion rate for different combinations
of $T_r \,\, \text{and}\,\, n $ as a function of the bath
temperature in the $\Phi$ dominated universe.}
\label{fig:hubble:FEU}
\end{figure}
In Fig.\,\,\ref{fig:hubble:FEU}, we show the expansion rate for different combinations of $T_r \,\, \text{and}\,\, n $ as a function of the bath temperature and compared it with the standard cosmology. In the left panel, we show the expansion rate for three different $n$ ($n=1$ for blue, $n=2$ for red, and $n=3$ for brown) by keeping $T_r$ fixed at 5 MeV. The green line corresponds to standard radiation dominated universe where $\rho_\Phi = 0$. We see that the expansion rate increases with increasing $n$ as it increases the total energy content of the universe. Similarly, the right panel shows the dependence on $T_r$, the temperature below which $\rho_{\Phi}$ becomes subdominant. One cane notice that smaller the $T_r$ longer the $\Phi$ domination and faster the expansion as shown in the right panel of the Fig.\,\,\ref{fig:hubble:FEU}. However, $T_r$ and $n$ are not completely independent parameters as the lower value of $T_r$ must be larger than $T_{BBN}\approx 1 \text{ MeV}$. For $T_r$ very close to the BBN temperature, the faster expansion of the universe can modify the prediction of the abundance of the light elements prior to BBN. As discussed in \cite{Luo:2020sho}, to avoid such effects we must satisfy the following condition,
\begin{equation}
 T_r  \geq \left(15.4 \right)^\frac{1}{n} \text{ MeV}\,.
\end{equation}

\begin{figure}[htb!]
\includegraphics[scale=0.53]{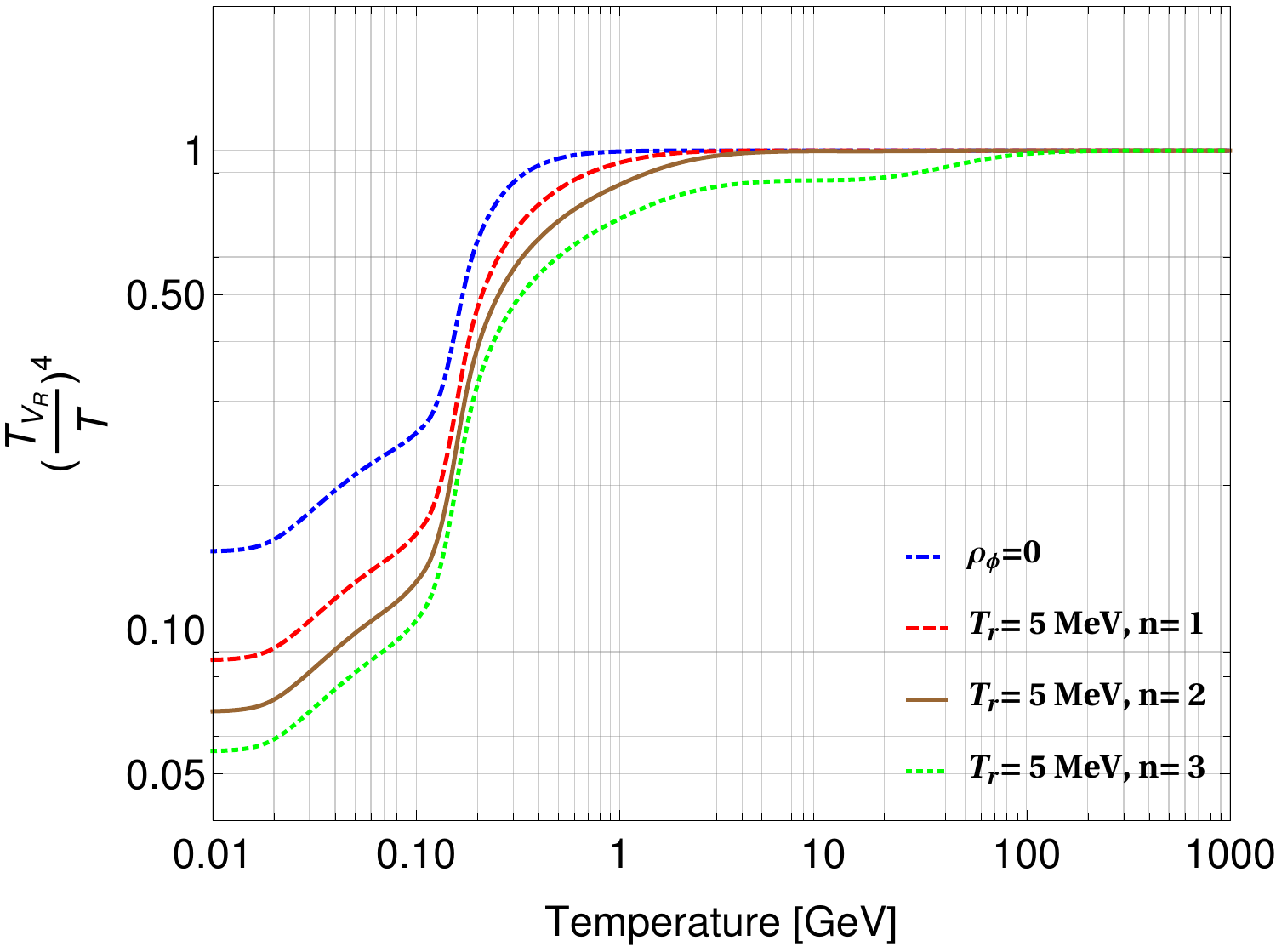}\,
\includegraphics[scale=0.53]{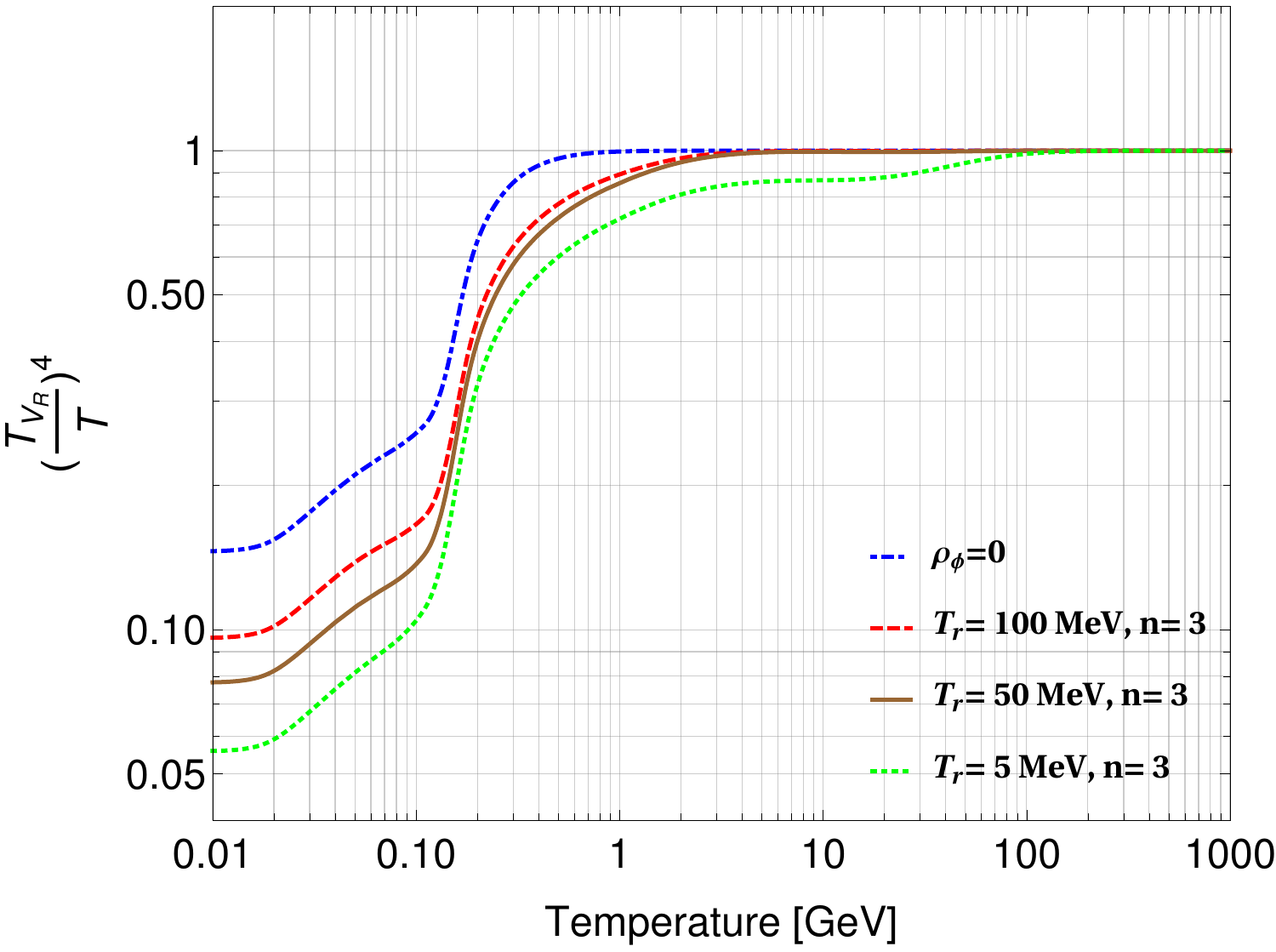}
\caption{Evolution of $T_{\nu_R}$ for different values of the parameters involved in the faster expansion of the universe for a given ${G_V=10^{-7}\, {\rm GeV}^{-2}}$.} 
\label{fig:dec:FEU}
\end{figure}
Let us now understand the impact of $T_r$ and $n$ on the decoupling temperature of the right-handed neutrinos $\left( \nu_{R}\right)$ which is shown in Fig.\,\,\ref{fig:dec:FEU}. From the above discussion, it is clear that both $T_r$ and $n$ can significantly increase the expansion rate of our universe and as a result affects the decoupling temperature of $\nu_R$. For a given interaction rate $\left(\Gamma\right)$, $\nu_R$ would decouple from the thermal bath at some higher temperature than the standard scenario (as discussed in \cite{Luo:2020sho}) because of the faster expansion. Due to their early decoupling, their final temperature also become smaller than the usual scenario and the contribution to $\Delta {\rm N_{eff}}$ would also decrease. In Fig.\,\,\ref{fig:dec:FEU}, we have shown the footprint of $T_r$ and $n$ in the final temperature of $\nu_R$. The left panel shows the dependence on $n$ while we keep $T_r$ fixed at 5 MeV and the right panel manifests the effect of $T_r$ by considering a fixed $n=3$. We have shown the evolution of $T_{\nu_R}$ in the standard radiation dominated universe by blue dot-dashed line. In both the cases, we set the effective coupling constant ${G}_{V}= {\rm 10^{-7}\, GeV^{-2}}$ ($8.57\times 10^{-3}\,G_F$). One very crucial point to note here is that ${G}_{V}= {\rm 10^{-7}\, GeV^{-2}}$ lies well above the upper limit given in Eq.\,\,\eqref{eqn:upp:lim}. This can also be understood by looking at the final value of $\left({ T_{\nu_R}/T}\right)^4 = 0.146$ which gives $\Delta {\rm N_{eff}} = 0.438$
(by using Eq.\,\,\eqref{eqn:Neff3}) where we consider the standard expansion history.
However, for sufficiently fast expansion $\nu_R$ can decouple at much higher temperature and resulting to smaller
$\Delta {\rm N_{eff}}$. For an example, if we set $T_r = 5$ MeV and $n=2$ the final value of 
$\left({T_{\nu_R}/T}\right)^4$ would be $0.0676$ as shown in the brown line
in the left panel of Fig.\,\,\ref{fig:dec:FEU}, corresponding to $\Delta {\rm N_{eff}} = 0.203$ which becomes allowed from PLANCK data \cite{Planck:2018vyg} at $2\sigma$ CL. So, we can reclaim the parameter space disfavored from the standard cosmology with the help of faster expansion of our universe. 
 
\begin{figure}[htb!]
\includegraphics[scale=0.42]{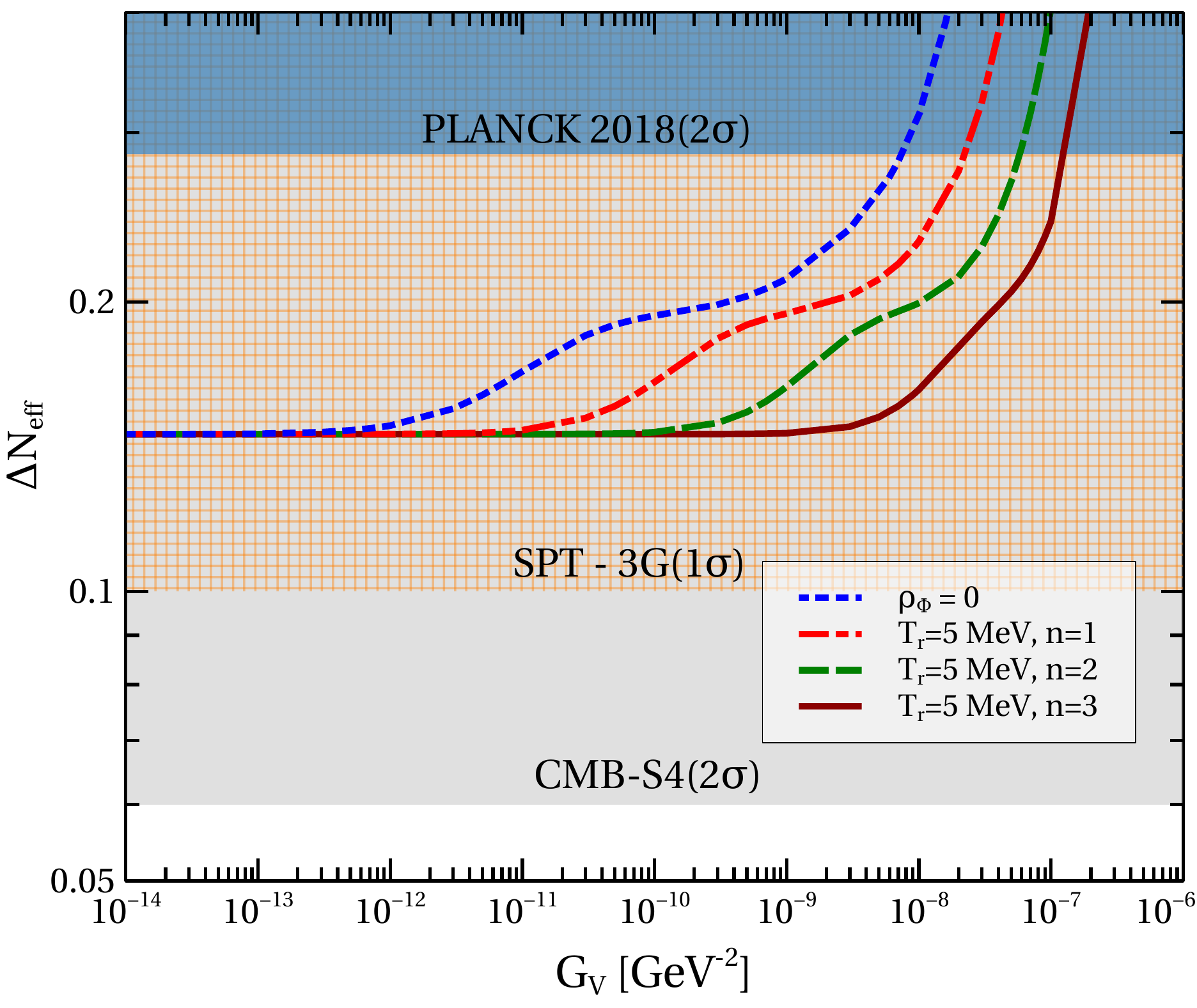}\,
\includegraphics[scale=0.42]{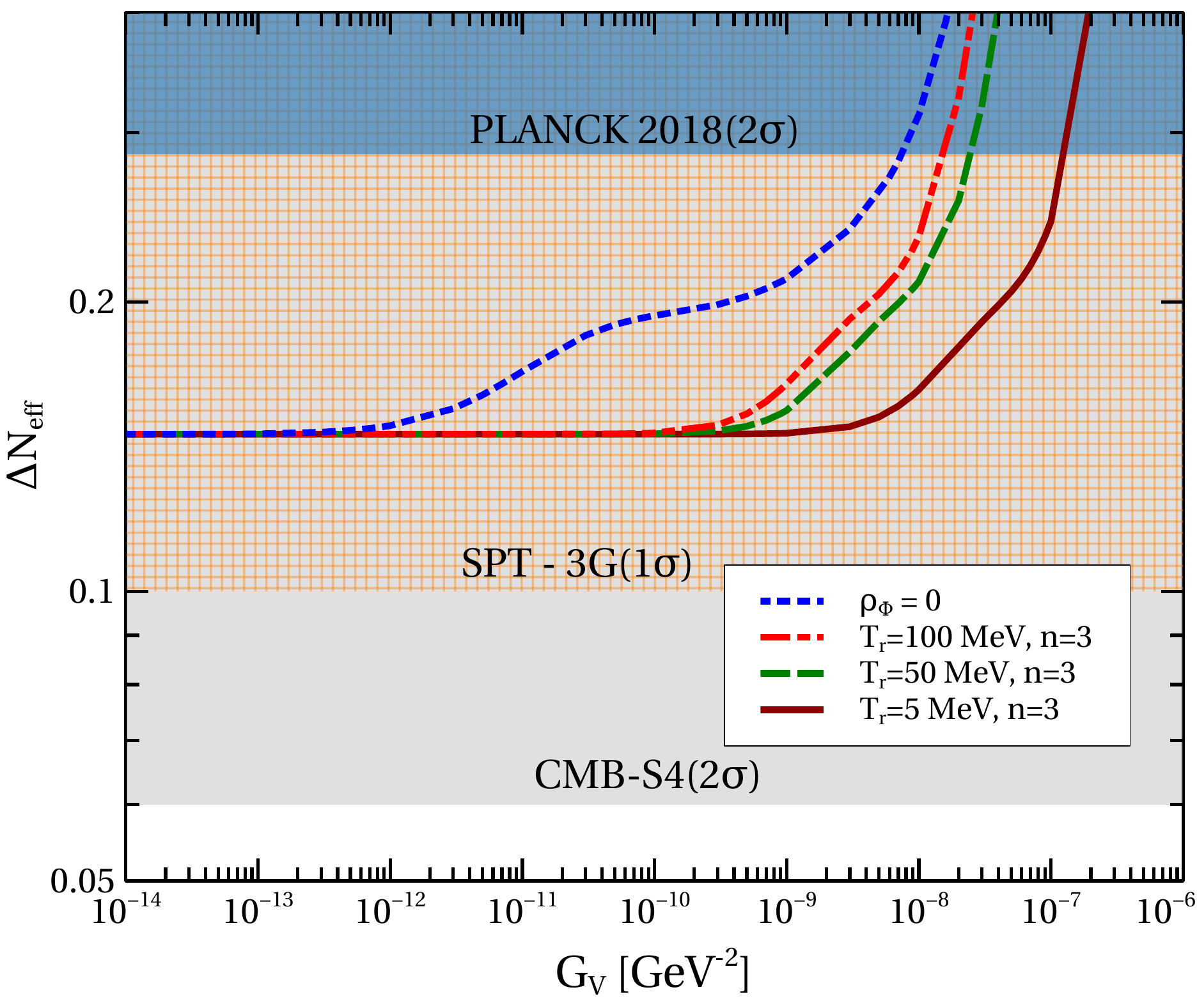}
\caption{The impact of $\nu_R$ to the effective relativistic degrees of freedom or ${\rm \Delta{N_{eff}}}$ in a early $\Phi$ dominated universe. Here, we have
considered interaction for $G_V$.} 
\label{fig:bound:FEU}
\end{figure}
Finally in Fig.\,\,\ref{fig:bound:FEU}, we present the main result of this work. We have shown that the upper limits given in Fig.\,\,\ref{fig:bound:standard} or in Eq.\,\,\eqref{eqn:upp:lim} will drastically change if we consider a non-standard cosmological evolution of the universe. The left panel is showing the impact of $n$ ${(n=1\,({\rm red}), n=2\,({\rm green})\,\,{\rm and}\,\,
n=3\,({\rm brown}))}$ while we have kept $T_r$ fixed at 5 MeV and the right panel manifests the effect of $T_r$ ${(T_r=100\,{\rm MeV\,(red)}, T_r=50\,{\rm MeV\,(green)}\,\,{\rm and}\,\,T_r=5\,{\rm MeV\,(brown)})}$ by considering a fixed $n=3$. The blue dashed line represents the ${\rm \Delta{N_{eff}}}$ in the standard radiation dominated universe where $\rho_\phi=0$. We have presented the results for $G_V$ ranging from ${\rm 10^{-14}\,\, GeV^{-2}}$ to ${\rm 10^{-6}\,\, GeV^{-2}}$ and set the other couplings to be zero. The behavior will be same for the other kind of interactions as well. In the Table \ref{tab:2B}, we have shown the changes of the upper bound of different kind of interactions as the ratio between the upper limit in the early $\Phi$ dominated universe and upper limit in the standard cosmology. notice that the more significant change is happening for large $n$ and small $T_r$ as this governs the faster expansion for a longer period of time.   

\begin{table}[h!]
\centering
\begin{tabular}{|l||l|l|l|l|}
 \hline
     \multirow{2}{*}{\ \ \ \ \ \ $T_r$, $n$} &
 \multicolumn{4}{|c|}{ Early $\Phi$ dominated universe}  \\
 \cline{2-5}

 & {\bf \hspace{0.3cm} $G_{S}^{\Phi}/G_{S}$} \hspace{0.3cm} & {\bf \hspace{0.3cm}${ G_{P}^\Phi/G_{P}}$\hspace{0.3cm}} & {\bf\hspace{0.3cm} ${G_{V}^{\Phi}/G_{V}}$\hspace{0.3cm}} &{\bf \hspace{0.3cm} ${G_{T}^{\Phi}/G_{T}}$\hspace{0.5cm}}\\
\hline
5 MeV, 1 & 2.98  & 3.01 & 3.02 & 2.98\\ 
\hline
5 MeV, 2 & 7.67  & 7.78 & 7.78 & 7.67\\ 
\hline
5 MeV, 3 & 18.45 & 18.79 & 18.79 & 18.45\\ 
\hline
50 MeV, 3 & 3.88 & 3.36 & 3.36 & 3.88\\ 
\hline
100 MeV, 3& 2.5 & 2.16 & 2.16 & 2.5 \\ 
\hline
\end{tabular}
\caption{Ratio of the upper bound on different types of interactions}
\label{tab:2B}
\end{table}

\section{C\lowercase{onclusion}}
\label{sec:concl}
The origin of small neutrino masses and the nature of neutrinos are
two unsolved puzzles of fundamental particle physics so far and
new particles and interactions are necessary to address these issues.
The Dirac neutrino framework is particularly interesting to study
as it contains at least two right-chiral components as light as the left-handed
ones, which can leave their signature in ${\rm N_{eff}}$
and can be probed in the current as well as future experiments
measuring cosmic microwave background photons.
However, this requires new non-standard interactions
to thermalise $\nu_R$ with the SM bath at the
early universe as the contribution to ${\rm N_{eff}}$ by Dirac neutrinos acquiring
masses only via the Standard Higgs mechanism (like other SM fermions) is
too small to be detected even in the next generation experiments also.  
Therefore, we have considered all possible types (e.g. scalar, pseudo scalar,
vector, tensor) of interactions between $\nu_L$ and $\nu_R$ in terms of
effective four-fermion operators. In order to calculate ${\rm \Delta{N_{eff}}}$
due to $\nu_R$, we need the temperature ratio $T_{\nu_R}/T$ just before
the decoupling of left-handed neutrinos. This we have obtained after solving
the Boltzmann equation for the most general collision term with FD statistics
and Pauli blocking factors. First, we have considered the standard cosmological
history and put upper bounds on the effective couplings of four-fermion
operators using the latest $2\sigma$ bound on ${\rm \Delta{N_{eff}}}$
by the Planck collaboration. The most stringent constraint is
obtained for the tensorial interaction between $\nu_L$ and $\nu_R$
where the upper limit on the coupling is $G_{T}<4.56\times 10^{-5}\,G_F$.
Our results have matched pretty well with earlier study in the literature. 

Next, we have considered non-standard cosmological evolution
of the universe prior to the BBN, where energy density of the
universe is dominated by some species with energy density
$\rho_i \propto a^{-(4+n)}$ and $n\neq0$. We have explored two different
cases i) $n=-1$ (early matter ($M$) domination) and ii) $n>0$ (early $\Phi$
domination, which is neither matter nor radiation). In the case of early $\Phi$
domination, we have only the effect of fast expansion before the BBN
as the Hubble parameter in this era  is $H\propto T^{2+n/2}$ ($n>0$). 
This leads to an early decoupling of $\nu_R$ from the thermal bath
which results in a reduced ${\rm \Delta{N_{eff}}}$. However, in this case,
the magnitude of ${\rm \Delta{N}_{eff}}$ due to $\nu_R$ is bounded from below.
That means, no matter how early the decoupling occurs,
there is always a minimum value of ${\rm \Delta{N}_{eff}}\sim 0.14$
for three right-handed neutrinos if they are in thermal bath at some epoch
and this minimum value is independent of the type of interaction
that thermalises $\nu_R$. The next generation CMB experiments
like CMB-S4, SPT-3G etc. with much improved sensitivity can easily validate
the idea of thermalised $\nu_R$ in the context of standard cosmology
and also for early $\Phi$ domination. 

On the other hand, the situation is different for the case with early matter domination. 
Like the universe with $\Phi$ domination, the matter dominated universe ($n=-1$)
also expands at a faster rate due to the presence of additional energy
in the form of non-relativistic matter over the standard radiation. This
also causes an early decoupling of $\nu_R$. However, unlike the previous
case, this is not the only thing that affects ${\rm \Delta{N}_{eff}}$.
In this case, the universe also undergoes a non-adiabatic expansion when
the matter field decays into the radiation composed of the SM particles. 
This results in a slowly cooling universe since
the entropy injection in the SM bath
alters the standard temperature-scale factor relation to $T\propto a^{-3/8}$.
Accordingly, the early decoupling of $\nu_R$ followed by slower
redshift of the SM bath temperature in contrast to the
$\Lambda$CDM cosmology, reduces the ratio $T_{\nu_R}/T$ at
$T\sim \mathcal{O}({\rm MeV})$ substantially.
Therefore, for an early matter dominated universe, 
the upcoming CMB experiments may not be able 
to trace $\nu_R$ in ${\rm N_{eff}}$ if there exists a
prolonged non-adiabatic phase ($\tau_M >>t_e$).
\section{A\lowercase{cknowledgements}}
Two of the authors AB and DN would like to thank Sougata Ganguly
for useful discussions and his help for computational purposes. 
The research work of AB is supported by Basic Science Research Program
through the National Research Foundation of Korea(NRF) funded by the Ministry
of Education through the Center for Quantum Spacetime (CQUeST) of
Sogang University (NRF-2020R1A6A1A03047877). The work of DN is partly
supported by National Research Foundation of Korea(NRF)’s grants,
grants no. NRF-2019R1A2C3005009(DN). 
\appendix
\section{The Boltzmann equations}
\label{app:BE}
In this appendix, we have described the Boltzmann equation expressing the
evolution of energy density ($\rho_{\nu_R}$) of $\nu_R$  with respect to
the photon temperature ($T$). In particular, we have focused on the collision
term for a general $\nu_L-\nu_R$ scattering taking into account the
Fermi-Dirac (FD) distribution functions for neutrinos and appropriate Pauli 
blocking factors. Following our earlier works \cite{Biswas:2021kio} on the
Dirac nature of neutrinos and its impact on $\Delta{\rm N_{eff}}$,
we can write the Boltzmann equation of $\rho_{\nu_R}$ for a
general $2\rightarrow2$ scattering as follows.
\begin{eqnarray}
\dfrac{d}{dt} \rho_{\nu_R}
+ 3 \mathcal{H} \left( \rho_{\nu_R} + 
P_{\nu_R}\right)
&=& \mathcal{C}_{2\rightarrow2}\,\,,
\label{eq:BE_rhonuR}
\end{eqnarray}
where $P_{\nu_R}$ is the pressure of $\nu_R$ which we have considered
equal to $\rho_{\nu_R}/3$ as in the case of radiation. The collision
term for a general $2\rightarrow2$ scattering like $\nu_i (P_1) + \nu_j (P_2) \leftrightarrow \nu_k (P_3) + \nu_l (P_4)$ is given by,
\begin{eqnarray}
\mathcal{C}_{2\rightarrow2} &=& \int d{\Pi_1}
d{\Pi_2}d{\Pi_3}d{\Pi_4}\left(2\pi\right)^4
\delta^4\left(P_1+P_2-P_3-P_4\right)
S\,|M|^2 \left(\Delta E_1 - \Delta^\prime E_3\right)
\Lambda(f_1,f_2,f_3,f_4)\,, \nonumber 
\label{eq:collision}\\
\end{eqnarray}
As we are interested in the species $\nu_R$, we are considering only
those $2\rightarrow2$ scatterings where in the initial state
we have at least one $\nu_R$ while the final state depending
on a particular process may or may not have $\nu_R$. The four
momentum of a species is denoted by $P_i$ and the corresponding
three momentum and energy are denoted by $\Vec{p_i}$ and
$E_i$ respectively. The phase space element
$d\Pi_i = \dfrac{d^3\vec{p_i}}{\left(2\pi\right)^3\,2\,E_i}$ and
$S$ is the symmetry factor which is equal to $\dfrac{1}{2!}$ for each pair
of identical species in the initial and final states respectively. Moreover,
the factors $\Delta$ and $\Delta^\prime$ represent number of $\nu_R$ in
the initial and the final state. The distribution functions including
the Pauli blocking factors are within the quantity $\Lambda$ which has
the following expression
\begin{eqnarray}
\Lambda(f_1, f_2, f_3, f_4) = f_3 f_4 \left(1-f_1\right)\left(1-f_2\right)
-f_1 f_2 \left(1-f_3\right)\left(1-f_4\right)\,,
\end{eqnarray}
with $f_i = \dfrac{1}{\exp(E_i/T_i)+1}$ where $T_i=T$, the
photon temperature, for the $\nu_L$ while the temperature
of $\nu_R$ is denoted by $T_{\nu_R}$. In Eq.\,(\ref{eq:collision}),
$|M|^2$ is the Lorentz invariant matrix amplitude square. We have
listed the expressions of $|M|^2$ in Table \ref{tab:modMsqr} for
all the relevant scattering processes involving $\nu_R$.

In order to proceed further, we need to simplify the collision term
with twelve dimensional integration over the three momenta of initial
and final state particles. For that we have followed the prescription given
in \cite{Hannestad:1995rs, Kreisch:2019yzn}. Here we have described
the procedure mentioning important expressions which we require to
simplify the collision term. We first perform the integration
over $\vec{p_4}$ and for that we use the identity
\begin{eqnarray}
\dfrac{d^3\vec{p_4}}{2\,E_4} = d^4{P_4}\,\delta(P^2_4)\,\Theta(P^0_4)\,,
\end{eqnarray}
where, for simplicity, we have neglected the tiny neutrino
masses as those are many orders of magnitude smaller than the typical
energy of the neutrinos. The time component of the four momentum
$P_4$ needs to be greater than zero and this has been ensured
by the Heaviside step function $\Theta(P^0_4)$. The integration
over $d^4{P_4}$ in Eq.\,\,(\ref{eq:collision})
is now done using four dimensional Dirac delta function. As a result,
we replace $P_4$ by $P_1+P_2-P_3$ and
$P^2_4 = P^2_1 + P^2_2 + P^2_3 + 2\left(P_1.P_2-P_1.P_3-P_2.P_3\right)$.
 Therefore, the collision now reduces to
\begin{eqnarray}
\mathcal{C}_{2\rightarrow2} &=& 2\,\pi\int d{\Pi_1}
d{\Pi_2}d{\Pi_3}\,
S\,|M|^2 \left(\Delta E_1 - \Delta^\prime E_3\right)\,
\delta\left(2\left(P_1.P_2-P_1.P_3-P_2.P_3\right)\right)
\times \nonumber \\
&&\,\,\,\,\,\,\,\,\,\,\,\,\,\,\,
\Theta\left(|\vec{p_1}+\vec{p_2}-\vec{p_3}|\right)
\Lambda(f_1,f_2,f_3,f_4(\vec{p_1}+\vec{p_2}-\vec{p_3}))\,. 
\label{eq:collision1}
\end{eqnarray}
Now, we need to choose a specific reference frame. We consider the vector $\vec{p_1}$ along the $Z$ axis. The vector $\vec{p_2}$ has polar
angle $\alpha$ and azimuthal angle $\beta$ while the vector $\vec{p_3}$
has polar angle $\theta$ and azimuthal angle $\phi$ respectively. However,
if we express all the dot products in Eq.\,\,(\ref{eq:collision1}) in terms of polar and azimuthal angles, we will find that only the difference between
the azimuthal angles of $\vec{p_2}$ and $\vec{p_3}$ is important. Hence,
in the subsequent calculations we consider $\beta$ as the difference
between two azimuthal angles. Therefore all the necessary scalar
products of three momenta are given by
\begin{eqnarray}
\vec{p_1}.\vec{p_2} &=& p_1 p_2 \cos\alpha\,, \nonumber\\
\vec{p_1}.\vec{p_3} &=& p_1 p_3 \cos\theta\,, \nonumber \\
\vec{p_2}.\vec{p_3} &=& p_2 p_3 \left(\sin\alpha \sin\theta\cos\beta 
+ \cos\alpha\cos\theta\right)\,.
\end{eqnarray}
 
 The next step is to do integration over the azimuthal angle $\beta$
 and the argument of the Dirac delta function is
 \begin{eqnarray}
 g(\beta) &=& 2\left(p_1 p_2 \left(1-\cos\alpha\right) -
 p_1 p_3 \left(1-\cos\theta\right) -
 p_2p_3\left(1-\cos\alpha\cos\theta\right)
 +p_2p_3\sin\alpha\sin\theta\cos\beta
 \right) \,, \nonumber \\
 {\rm and}&&\\
 &&~~~~~~~~~~~~~~~~~~~~~~~~~\dfrac{dg(\beta)}{d\beta} =
 -2p_2p_3 \sin\alpha\sin\theta\sin\beta \,.
 \label{eq:gbeta}
 \end{eqnarray}
Using the well known property of the Dirac delta function,
\begin{eqnarray}
\bigintsss d\beta \delta(g(\beta)) = \sum_i \bigintss d\beta
\dfrac{\delta(\beta-\beta_i)}
{~~~~\left|\frac{dg(\beta)}{d\beta}\right|_{\beta=\beta_0^i}}
\,,
\end{eqnarray}
the integration over $\beta$ can be done easily where the roots
$\beta_0$s can be obtained by setting $g(\beta) = 0$ which gives
\begin{eqnarray}
\cos \beta_0 = -\dfrac{p_1 p_2 \left(1-\cos\alpha\right) -
 p_1 p_3 \left(1-\cos\theta\right) -
 p_2p_3\left(1-\cos\alpha\cos\theta\right)}
 {p_2p_3\sin\alpha\sin\theta}\,,
 \label{eq:cosbeta}
\end{eqnarray}
and this results in two different values of $\beta_0$
lying between [0,$\pi$] and [$\pi$, 2$\pi$] for one
particular value of $\cos \beta_0$. The natural condition
$\cos^2{\beta_0}\leq 1$ automatically sets $\sin^2{\beta_0}\geq 0$
and it further implies from Eq.\,\,(\ref{eq:gbeta})
that $\left|\dfrac{dg(\beta)}{d\beta}\right|^2_{\beta=\beta_0}\geq 0$.
As the integrand is symmetric in $\beta$, we can write
\begin{eqnarray}
\bigintsss_0^{2\pi} d\beta \delta(g(\beta)) = 2\,\bigintss_0^{\pi} d\beta
\dfrac{\delta(\beta-\beta_0)}
{~~~~\left|\frac{dg(\beta)}{d\beta}\right|_{\beta=\beta_0}}
\Theta\left(\left|\dfrac{dg(\beta)}{d\beta}\right|^2_{\beta=\beta_0}\right)
\,.
\end{eqnarray}
Substituting $\sin{\beta_0}$ using the expression of $\cos\beta_0$
(Eq.\,\,(\ref{eq:cosbeta})) in Eq.\,\,(\ref{eq:gbeta}), we can express
$\left|\dfrac{dg(\beta)}{d\beta}\right|_{\beta=\beta_0}$ in the
form of a quadratic equation in $\cos\alpha$ as
\begin{eqnarray}
&&\left|\dfrac{dg(\beta)}{d\beta}\right|_{\beta=\beta_0}=
2\sqrt{a_{\theta} \cos^2{\alpha} + b_{\theta} \cos{\alpha} +
c_{\theta}}\,\,, 
\label{eq:dgdbeta}\\
&&\hspace{-4cm}{\rm with} \nonumber \\
&& a_{\theta} = -p_2^2\left(p^2_1 + p_3^2 - 2p_1p_3\cos{\theta}\right)\,, 
\label{atheta} \\
&& b_{\theta} = 2p_2\left(p_1-p_3\cos\theta\right)
\left(p_1p_2-p_3\left(p_1+p_2\right) + p_1p_3\cos\theta\right)\,,
\label{btheta}\\
&& c_{\theta} = -\left(p_1p_2-p_3\left(p_1+p_2\right)+
p_1p_3\cos\theta\right)^2 + p^2_2p^2_3\left(1-\cos^2\theta\right)\,,
\label{ctheta}
\end{eqnarray}
After integrating over the azimuthal angle $\beta$, the collision term reduces to
\begin{eqnarray}
\mathcal{C}_{2\rightarrow2} &=& 
\dfrac{\left(2\pi\right)^2S}{2^3\left(2\pi\right)^9}\,
\int \dfrac{p^2_1 dp_1}{E_1}
\dfrac{p^2_2 dp_2}{E_2} \dfrac{p^2_3 dp_3}{E_3}\,
\dfrac{d(\cos\alpha) d(\cos\theta)}
{\sqrt{a_{\theta} \cos^2{\alpha} + b_{\theta} \cos{\alpha} +
c_{\theta}}}
|M|^2 \left(\Delta E_1 - \Delta^\prime E_3\right)\,
\times \nonumber \\
&&\,\,\,\,\,\,\,\,\,\,\,\,\,\,\,
\Theta\left(|\vec{p_1}+\vec{p_2}-\vec{p_3}|\right)
\Theta\left(a_{\theta} \cos^2{\alpha} + b_{\theta} \cos{\alpha} + 
c_{\theta}\right)
\Lambda(f_1,f_2,f_3,f_4(\vec{p_1}+\vec{p_2}-\vec{p_3}))\,, \nonumber \\ 
\label{eq:collision2}
\end{eqnarray}
where we have trivially integrated over the other azimuthal angle
$\phi$ also. The next step is to integrate over $\cos\alpha$ and
for that one needs the actual expression of $\left|M\right|^2$. However,
since $\left|M\right|^2$ is Lorentz invariant, we should only have
scalar products of four momenta and the polar angles will
enter in $\left|M\right|^2$ through $P_1.P_2$ and $P_1.P_3$ only\footnote{We
can replace the scalar product $P_2.P_3$ using the relation
$P_2.P_3 = P_1.P_2 - P_1.P_3$ for $m_{\nu} = 0$}. Therefore, in general,
the matrix amplitude square can be expressed a quadratic equation of 
$\cos{\alpha}$ whose coefficients depend on three momenta $p_1$, $p_2$,
$p_3$ and also on the other polar angle $\theta$ similar to
Eqs.\,\,(\ref{atheta}-\ref{ctheta}). The integration over $\cos{\alpha}$
can be done using the following results \cite{Kreisch:2019yzn},
\begin{eqnarray}
\bigintss_{-\infty}^{+\infty} \dfrac{dx}
{\sqrt{a_{\theta} x^2 + b_{\theta} x + c_{\theta}}}
\Theta\left(a_{\theta} x^2 + b_{\theta} x + c_{\theta}\right) &=& \dfrac{\pi}{\sqrt{-a_{\theta}}}
\Theta\left(b_{\theta}^2-4a_{\theta}c_{\theta}\right) \,\,,\\
\bigintss_{-\infty}^{+\infty} \dfrac{x\,dx}
{\sqrt{a_{\theta} x^2 + b_{\theta} x + c_{\theta}}}
\Theta\left(a_{\theta} x^2 + b_{\theta} x + c_{\theta}\right) &=&
\dfrac{b_{\theta}\,\pi}{2a_{\theta}\sqrt{-a_{\theta}}}
\Theta\left(b_{\theta}^2-4a_{\theta}c_{\theta}\right)\,\,,\\
\bigintss_{-\infty}^{+\infty} \dfrac{x^2\,dx}
{\sqrt{a_{\theta} x^2 + b_{\theta} x + c_{\theta}}}
\Theta\left(a_{\theta} x^2 + b_{\theta} x + c_{\theta}\right) &=& 
\dfrac{\pi\left(3b_{\theta}^2-4a_{\theta}c_{\theta}\right)}
{8a^2_{\theta}\sqrt{-a_{\theta}}}\,\,
\Theta\left(b_{\theta}^2-4a_{\theta}c_{\theta}\right)\,,
\end{eqnarray}
where, $x=\cos{\alpha}$. Although, the limit of the integrals
is between $[-\infty, +\infty]$, the actual limit is between
the two roots $\cos\alpha_{\pm} = \dfrac{-b_{\theta} \mp
\sqrt{b^2_{\theta}-4a_{\theta}c_{\theta}}}{2\,a_{\theta}}$
of the quadratic equation $a_{\theta}
\cos{\alpha}^2 + b_{\theta} \cos{\alpha} +c_{\theta}$ due to the
Heaviside step function. The roots will be bounded
between $[-1,1]$ when the discriminant
$b^2_{\theta}-4a_{\theta}c_{\theta}>0$. This will also determine the
actual limit of the integral over $\cos{\theta}$ between
$[\cos\theta_{-},1]$ where
\begin{eqnarray}
\cos\theta_{-} = \text{Max}\left[-1, \dfrac{p_3\left(p_1+2\,p_2\right)-2\,p_2
\left(p_1+p_2\right)}{p_1p_3}\right]\,\,{\rm for}\,\,p_1+p_2\geq p_3\,.
\label{costhm}
\end{eqnarray}
Finally, the collision term with integrations\footnote{We
have numerically computed the multi-dimensional integration
using \href{http://www.feynarts.de/cuba/}{Cuba library} \cite{Hahn:2004fe}.} 
over $p_1$, $p_2$, $p_3$ and two polar angles is given by
\begin{eqnarray}
\hspace{-1cm}
\mathcal{C}_{2\rightarrow2} &=& 
\dfrac{S}{2^3\left(2\pi\right)^7}\,
\int_0^{\infty} \dfrac{p^2_1 dp_1}{E_1}
\int_0^{\infty} \dfrac{p^2_2 dp_2}{E_2}
\int_0^{\infty} \dfrac{p^2_3 dp_3}{E_3}\,
\int_{\cos{\theta{-}}}^{1} d(\cos\theta)
\int_{\cos{\alpha{+}}}^{\cos\alpha_{+}}
\dfrac{d(\cos\alpha)\,\,|M|^2}
{\sqrt{a_{\theta} \cos^2{\alpha} + b_{\theta} \cos{\alpha} +
c_{\theta}}}
\times \nonumber \\
&&\,\,
 \left(\Delta E_1 - \Delta^\prime E_3\right)
\Theta\left(|\vec{p_1}+\vec{p_2}-\vec{p_3}|\right)
\Theta\left(a_{\theta} \cos^2{\alpha} + b_{\theta} \cos{\alpha} + 
c_{\theta}\right)
\Lambda(f_1,f_2,f_3,f_4(\vec{p_1}+\vec{p_2}-\vec{p_3}))\,, \nonumber \\ 
\label{eq:collision_term_final}
\end{eqnarray}
with $E_i=p_i$ in the present case with $m_{\nu} = 0$. For general case
when the masses of the initial and final state particles cannot be
neglected, the expressions given in Eqs.\,(\eqref{eq:cosbeta},
\eqref{atheta}-\eqref{ctheta} and \eqref{costhm}) will be different
and can be found in \cite{Hannestad:1995rs}.

Now, substituting $\rho_{\nu_R} = \kappa T^4_{\nu_R}$ with
$\kappa = g_{\nu_R}\frac{7}{8} \frac{\pi^2}{30}$ and $g_{\nu_R} = 1$,\footnote{Since we have considered the contribution of
$\nu_R$ only in $\rho_{\nu_R}$. If we want to include
the effect of $\overline{\nu_R}$ in the energy density
(i.e. $g_{\nu_R} = 2$), we have to add a collision term 
for $\overline{\nu_R}$ also in the right hand side of
Eq.\,\eqref{eq:BE_rhonuR}. As the collision terms for $\nu_R$
and $\overline{\nu_R}$ are identical, this results in an
extra factor of 2 in the right hand side of Eq.\,\eqref{eq:BE_rhonuR}
which eventually is canceled out by $g_{\nu_R}$ within the
quantity $\kappa$ in the denominator.}
and transforming time $t$ by the photon temperature $T$
in Eq.\,\eqref{eq:BE_rhonuR}, we get the evolution equation for
$\xi=T_{\nu_R}/T$ with $x=M_0/T$ as given below
\begin{eqnarray}
x\dfrac{d\xi}{dx} + \left(\beta -1\right) \xi =
\dfrac{\beta\,x^4}{4\,\kappa\,\xi^3{H}M_0^4} 
\mathcal{C}_{2\rightarrow2}\,,
\label{eq:BE_xi}
\end{eqnarray}
where $M_0$ is any arbitrary mass scale, and
$\beta = \left(1 -\dfrac{1}{3}\dfrac{x}{g_s}\dfrac{dg_s}{dx}\right)$
with $g_s$ is the number of degrees of freedom associated with
the entropy density.


\bibliographystyle{JHEP}
\bibliography{ref.bib} 
\end{document}